\documentclass[12pt,a4paper]{article}

\usepackage{latexsym}
\usepackage{setspace}
\usepackage{a4wide}
\usepackage{amsmath}
\usepackage{amsfonts}
\usepackage{amscd}
\usepackage{amssymb}
\usepackage{amsbsy}
\usepackage{mathrsfs}
\usepackage{yfonts}
\usepackage{bbm}
\usepackage[dvips]{epsfig,graphics}
\usepackage{epsfig}
\usepackage{graphicx}
\usepackage{psfrag}
\usepackage{upgreek}
\usepackage{undertilde}
\usepackage{color}
\usepackage{leftidx}
\usepackage{mathbbol}

\DeclareMathAlphabet{\mathpzc}{OT1}{pzc}{m}{it}

\begin{document}

{\color{red}{This document is the Accepted Manuscript version of a Published Work that appeared in final form in International Journal of Nonlinear Mechanics, copyright Elsevier after peer review and technical editing by the publisher. To access the final edited and published work see \\ http://www.sciencedirect.com/science/article/pii/S0020746214001826}}

\title{Curvature dependent surface energy for a free standing monolayer graphene: some closed form solutions of the nonlinear theory}

\author{D. Sfyris, G.I. Sfyris \& C. Galiotis}

\maketitle

\begin{abstract}

Continuum modeling of a free-standing graphene monolayer, viewed as a two dimensional 2-lattice, requires specifications of the components of the shift vector that act as an auxiliary variable. The field equations are then the equations ruling the shift vector, together with momentum and moment of momentum equations. We present an analysis of simple loading histories such as axial, biaxial tension/compression and simple shear for a range of problems of increasing difficulty. We start by laying down the equations of a simplified model which can still capture bending effects. Initially, we ignore out of plane deformations. For this case, we solve analytically the equations ruling the auxiliary variables in terms of the shift vector; these equations are algebraic when the loading is specified. As a next step, still working on the simplified model, out-of-plane deformations are taken into account and the equations complicate dramatically. We describe how wrinkling/buckling can be introduced into the model and apply the Cauchy-Kowalevski theorem to get existence and uniqueness in terms of the shift vector for some characteristic cases. Finally, for the treatment of the most general problem, we classify the equations of momentum and give conditions for the Cauchy-Kowalevski theorem to apply.

\end{abstract}


\section{Introduction}

Graphene is a two dimensional sheet that constitutes the building unit of all graphitic forms of matter, such as graphite, carbon nanotubes and carbon fibers. For modeling graphene many different approaches at different scales can be found in the literature ranging from first principle calculations (\cite{Kudinetal2001,Liuetal2007}), atomistic calculations (\cite{Zakharchenkoetal2009,Zhaoetal2009}) and continuum mechanics (\cite{Cadelanoetal2009,Yakobsonetal1996,Lu-Huang2009}). Furthermore, mixed atomistic formulations with finite elements are being reported for graphene (\cite{Arroyo-Belytschko2002,Arroyo-Belytschko2004,Theodosiou-Saravanos2013}) based on the earlier notion of a quasi-continuum (\cite{Tadmoretal1999,Tadmoretal1996}). 

The mathematical theory of surface elasticity is established by Gurtin and Murdoch (\cite{Gurtin-Murdoch1975}). This pure membrane approach is incapable of taking into account out of plane deformations. Generalization of this framework to take into account bending effects is given by Steigmann and Ogden (\cite{Steigmann-Ogden1999}). These authors propose a surface energy which depends, apart from a surface measure of the deformation, on the curvature tensor as well, in similar trends with previous works (\cite{Cohen-DeSilva1966,Murdoch-Cohen1979}). The curvature tensor is a measure of the out-of-plane deformations the surface suffer and this way bending effects are introduced into the framework. Steigmann and Ogden, in the same work (\cite{Steigmann-Ogden1999}), also describe a rigorous way for defining the notion of material symmetry for curvature dependent surface energies in line with Noll's fundamental work (\cite{Noll1958}). Implications of such energies for nanostructures are studied by Chappadia etal. (\cite{Chhapadiaetal2011}). 

In a recent work (\cite{Sfyris-GaliotisMMS}) we adopt the framework of Steigmann and Ogden (\cite{Steigmann-Ogden1999}) and utilize a surface energy function depending on three arguments for a free standing monolayer graphene. The first one is an in-surface strain measure describing changes happening on the surface. The second argument is the curvature tensor which describes the out of surface motions and introduces bending effects into the model. The third argument is the shift vector (SV) which connects the two simple lattices when graphene is seen as a monoatomic 2-lattice. The motivation for assuming the shift vector as an independent variable comes from the work of Pitteri and Zanzotto (\cite{Pitteri-Zanzotto2003} and references therein). These authors utilize an energy function depending on the shift vector when modeling a multilattice. We note that for graphene a similar assumption is made by E and Ming (\cite{E-Ming2007}). 

Using the above surface energy, calculation of the surface stress and the surface couple stress tensor at the continuum level is possible. This way the number of independent relations to be observed in experiments becomes available; these are 13 independent material parameters, in the simplest expression of the model. The surface stress tensor is responsible for in-plane motions while out-of-plane motions are due to the surface couple stress tensor. The elasticities of the material can then be calculated and one may also lay down the field equations characterizing the problem: the momentum, the moment of momentum equation as well as an equation for the evaluation of the shift vector.  

Being aware of the molecular theories of elasticity, where the energy depends on the lattice vectors, we stress that this analysis (\cite{Sfyris-GaliotisMMS}) is confined to weak transformation neighborhoods (\cite{Pitteri1984}). This way the classical theory of invariants for continuum mechanics can be utilized, so we can obtain the invariants of the surface energy function. This is compatible with molecular theories when the Cauchy-Born rule is enforced, and also compatible with the global theory of Ericksen (\cite{Ericksen2008,Fosdick-Hertog1990,Ericksen1979}). 

In this respect, we present the key findings of \cite{Sfyris-GaliotisMMS} which is the theoretical background for this work. When graphene is viewed as a monoatomic 2-lattice, its arithmetic symmetries can be deduced from the fundamental work of Fadda and Zanzotto (\cite{Fadda-Zanzotto2000}). To arrive at the classical symmetries, those employed by continuum mechanics, the analysis should be confined to weak transformation neighborhoods (\cite{Pitteri-Zanzotto2003,Pitteri1984,Pitteri1985}). Also the Cauchy-Born rule (\cite{Ericksen2008}) should be enforced. Under these assumptions, we work at the continuum level with an energy depending on three arguments: an in-surface strain measure, the curvature tensor and the shift vector. Since symmetries are now those employed by continuum mechanics, we are able to deduce the complete and irreducible representation of graphene's energy. This way calculation of the surface stress and the surface couple stress tensor becomes possible. These tensors participate to the field equations ruling the problem: the momentum, the moment of momentum equation and the shift vector. In Section 3 we derive the field equations in terms of the kinematic variables: the position vector of the points of the surface, the components of the curvature tensor and the components of the shift vector. These equations are designed for the geometrical and materially nonlinear case. 

The need for describing graphene using nonlinear elasticity is based on graphene's very high strength. Efficient computational methods, such as ab-initio and/or molecular mechanics, report that graphene can deform elastically at tension up to more than 20 per cent of strain (see e.g. \cite{Liuetal2007}). Compression can also reach such high levels, even thought buckling occurs at lower strains; this buckling is elastic so graphene can accommodate even higher compressive strains in an elastic manner (see e.g. \cite{Zhuetal2012}). The present approach is designed as the theoretical backbone, at the continuum level, of this nonlinear behaviour graphene shows.   

Earlier attempts to use nonlinear elasticity for graphene can be found to the work of Lee etal. (\cite{Leetal2008}) who use a nanoidentation experiment in an atomic force microscope to measure the elastic properties and intrinsic strength of graphene. Using second order elasticity they evaluate Youngs modulus, the second order elastic constant as well as graphene's breaking strength. Their analysis models graphene as an isotropic body in one dimension, due to symmetry in the loading. Generalization of their approach to two dimensions is done by Cadelano et al. (\cite{Cadelanoetal2009}). These authors view graphene as an isotropic body and they utilize an energy cubic in strains (second order elasticity in words of Murnaghan (\cite{Murnaghan1951}) and Rivlin \cite{Rivlin1963}). Utilizing tight-binding atomistic simulations they calculate Young's modulus, Poisson ratio as well as higher order constants for graphene. While interesting and novel their approach is, it lacks the treatment of bending effects. It also models graphene as an isotropic body; dependence on the zigzag and the armchair direction is not incorporated to the constitutive law through dependence on a structural tensor. Fifth order models for graphene are presented by Wei et al. (\cite{Weietal2009}). These authors utilize an energy that depends on strains of the fifth order. Using density functional theory for simple loading histories they evaluate higher order constants for graphene. Their approach does not include bending effects neither anisotropy; graphene is modeled as an isotropic body.

At Section 4, a model for the problem is presented, where five (5) out of the thirteen (13) material parameters of the model are set to zero, with the purpose of simplifying the mathematical analysis while capturing bending effects. Initially, by focusing on in-plane motions for simple mechanical loadings, we disregard dependence on the curvature tensor. As an outcome of that, the equation of moment of momentum need not be taken into account. By also assuming that the shape of the body, at the reference state, is a rectangular plate, we examine axial, biaxial tension/compression and simple shear loadings. The strategy consists of assuming the form of the solution for the position vector $\bf x$ of the surface and seeking for suitable forms of the SV that guarantees fulfillment of the field equations. The outcome consists of expressions for the SV, which is denoted by $\bf p$ that, in general, depend on the material parameters and the loading constant as well. What allows us to give these closed form solutions is the fact that the equations ruling the auxiliary variables are algebraic and solvable in terms of $\bf p$.  

When out-of-plane motions are taken into account, the field equations of the simplified model become much more complicated. In our model wrinkling/buckling is a product of in-plane mechanical tension/compression. The equations ruling the auxiliary variables are algebraic as previously, but now they are not solvable in a closed form. We describe how wrinkling/buckling can be introduced into the present framework following standard assumptions on the topic (\cite{Timoshenko,Punteletal2011}) and write down the field equations describing the problem at hand. More specifically, we treat the case when wrinkling/buckling is a product of tension/compression on the in plane. We note that the different behaviour of graphene at tension and compression is not taken into account here, since that would require extension of the model to include this hardening behaviour which is beyond the scope of this work. We classify the momentum equation viewed as a system of quasilinear equations for the shift vector and also give conditions for the Cauchy-Kowalevski theorem to apply. This theorem guarantees existence and uniqueness of solutions for the SV; these conditions are expressions in terms of the material parameters and the SV. The presence of the shift vector in such equations results from the fact that momentum equation is a quasilinear system in terms of $\bf p$. These are the contents of Section 5. 

In its most general form, the problem of free standing monolayer graphene sheet under mechanical loading is extremely difficult to tackle analytically; nevertheless, at Section 6 we present and classify the momentum equations and we also give necessary conditions for the Cauchy-Kowalevski theorem to apply. Finally, in Section 7 we conclude with a summary of the results and some remarks highlighting future directions. The appendix section giver a short reminder of issues like classification, existence and uniqueness of solutions for quasilinear systems (\cite{Ockendonetal}).

\section{Curvature dependent surface energy for graphene }
Following the  classification of 2-lattices by Fadda and Zanzotto (\cite{Fadda-Zanzotto2000}), we treat a monolayer graphene as a hexagonal monoatomic 2-lattice with unit cell of the form of Figure 1. 
\begin{figure}[!htb]
\centering
\includegraphics{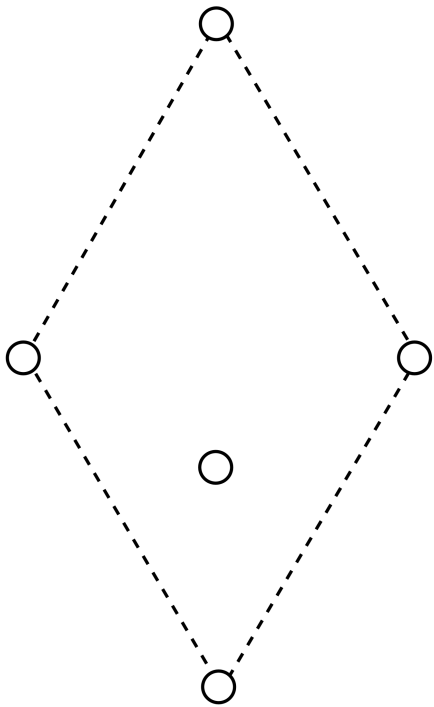}
\caption{The unit cell of a hexagonal 2-lattice (\cite{Fadda-Zanzotto2000}). }
\label{fig:digraph}
\end{figure}
The lattice and shift vectors are depicted in Figure 2
\begin{figure}[!htb]
\centering
\includegraphics{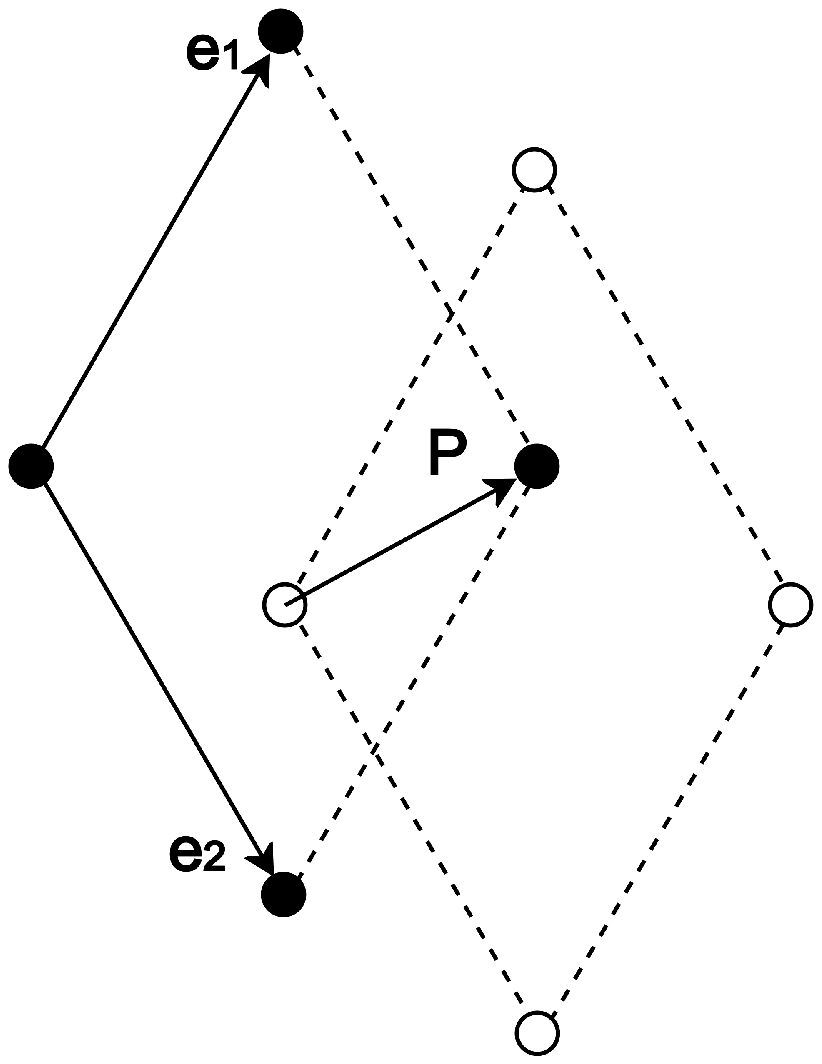}
\caption{The lattice and shift vectors of graphene. }
\label{fig:digraph}
\end{figure}
and defined as 
\begin{equation}
{\bf e}_1=(\sqrt{3} l, 0), \ \ {\bf e}_2=\left( \frac{\sqrt{3}}{2} l, \frac{3}{2} l \right), \ \ {\bf p}=\left( \frac{\sqrt{3}}{2} l, \frac{1}{2} l \right),
\end{equation}
$l$ being the lattice size, namely the interatomic distance at ease which is approximately 1, 42 Angstrom. The two simple hexagonal lattices are 
\begin{eqnarray}
&&L_1 (l)=\{ {\bf x} \in \mathcal R^2: {\bf x}=n^1 {\bf e}_1 + n^2 {\bf e}_2, \ \ (n^1, n^2) \in \mathcal Z^2 \}, \nonumber\\
&&L_2 (l)={\bf p}+L_1(l). 
\end{eqnarray}

The arithmetic symmetry group (\cite{Ericksen1979,Pitteri-Zanzotto2003}) of graphene is then described by the matrices 
\begin{equation}
\begin{pmatrix}
      -1 & -1 & -1 \\
      1 & 0 & 0 \\
      0 & 0 & 1
\end{pmatrix},
\begin{pmatrix}
      0 & 1 & 0 \\
      1 & 0 & 0 \\
      0 & 0 & 1
\end{pmatrix},
\begin{pmatrix}
      -1 & -1 & -1 \\
      0 & 1 & 0 \\
      0 & 0 & 1
\end{pmatrix},
\end{equation}
\begin{equation}
\begin{pmatrix}
      1 & 0 & 0 \\
      -1 & -1 & -1 \\
      0 & 0 & 1
\end{pmatrix},
\begin{pmatrix}
      1 & 0 & 0 \\
      0 & 1 & 0 \\
      0 & 0 & 1
\end{pmatrix},
\begin{pmatrix}
      0 & 1 & 0 \\
      -1 & -1 & -1 \\
      0 & 0 & 1
\end{pmatrix}.
\end{equation}
The eigenvalues of these matrices are $1, -1, e^{i \pi/3}, e^{-i \pi/3}$, so they describe the identity transformation, reflection transformation, and rotations by $60^0$, $-60^0$, respectively. 

At the continuum level, topologically, graphene is modeled as a two dimensional smooth surface embedded in a three dimensional Euclidean space. Position vectors on the reference configuration $\mathcal B_R$ of the referential surface are parametrized by two surface coordinates $\Theta^{\alpha}, \alpha=1, 2$ as (\cite{Ciarlet2005})
\begin{equation}
{\bf X}={\bf X}(\Theta^{\alpha}). 
\end{equation}  
After the deformation the surface occupies the current configuration $\mathcal B_C$, described by the position vector
\begin{equation}
{\bf x}={\bf x}(\Theta^{\alpha}). 
\end{equation} 
Covariant surface base vectors are then defined as 
\begin{equation}
{\bf A}_{\alpha}={\bf X}_{, \alpha }, \ \  {\boldsymbol \alpha}_{\alpha}={\bf x}_{, \alpha },
\end{equation}
for $\mathcal B_R$ and $\mathcal B_C$, respectively. Contravariant base vectors are given as
\begin{equation}
{\bf A}_{\alpha} \cdot {\bf A}^{\beta}=\delta_{\alpha}^{\beta}, \ \ {\boldsymbol \alpha}_{\alpha} \cdot {\boldsymbol \alpha}^{\beta}=\delta_{\alpha}^{\beta},
\end{equation}
$\delta_{\alpha}^{\beta}$ being the two dimensional Kronecker delta. 

The surface deformation gradient ${\bf F}_S$ reads
\begin{equation}
{\bf F}_S={\boldsymbol \alpha}_{\alpha} \otimes {\bf A}^{\alpha},
\end{equation}
while the right Cauchy-Green deformation tensor takes the form
\begin{equation}
{\bf C}_S={\bf F}_S^T \cdot {\bf F}_S.
\end{equation}
Being symmetric, the three components of ${\bf C}_S$ are given as
\begin{eqnarray}
&& C_{11}={\bf x}_{,1} \cdot {\bf x}_{,1}=x^2_{1,1}+x^2_{2,1}+x^2_{3,1} \\
&& C_{12}={\bf x}_{,1} \cdot {\bf x}_{,2}=x_{1,1}x_{1,2}+x_{2,1}x_{2,2}+x_{3,1}x_{3,2}=C_{21} \\
&& C_{22}={\bf x}_{,2} \cdot {\bf x}_{,2}=x^2_{1,2}+x^2_{2,2}+x^2_{3,2} 
\end{eqnarray}
The surface tensor ${\bf C}_S$ provides a frame indifferent measure of the in-plane deformation of the surface. 

Out-of-plane deformations are described by the surface curvature tensors
\begin{eqnarray}
&& {\bf b}_0=b_{0_{\alpha \beta}} {\bf A}^{\alpha} \otimes {\bf A}^{\beta}, \\
&& {\bf b}=b_{\alpha \beta} {\boldsymbol \alpha}^{\alpha} \otimes {\boldsymbol \alpha}^{\beta}, 
\end{eqnarray}
defined with respect to $\mathcal B_R$ and $\mathcal B_C$, respectively. These measures constitute the second fundamental form of the surfaces $\mathcal B_R$ and $\mathcal B_C$, respectively. Taking into account bending effects for a monolayer graphene modeled as a surface, requires dependence of the energy on the curvature (\cite{Steigmann-Ogden1999,Murdoch-Cohen1979,Cohen-DeSilva1966}). Thus for a monolayer graphene at the continuum level we assume an energy of the form (\cite{Sfyris-GaliotisMMS})
\begin{equation}
W=W({\bf C}_S, {\bf b}_0, {\bf p}). 
\end{equation}   
Dependence on the shift vector, $\bf p$, at the continuum level, results from the fact that at the crystalline level graphene is a 2-lattice. Now, we confine ourselves to weak transformation neighborhoods (\cite{Pitteri-Zanzotto2003}) and assume validity of the Cauchy-Born rule (\cite{Ericksen2008}). With these assumptions enforced we may utilize the classical symmetries employed by continuum mechanics. Following (\cite{Zheng1994,Zheng1993}) we use the symmetry group $\mathcal D_{6h}$ with generators ${\bf R}(\frac{2 \pi}{6}), {\bf R}_j$, in line with the eigenvalues of the matrices of eqs. (3), (4) for graphene.

Since $\mathcal D_{6h}$ is not the full isotropy group we use the principle of isotropy of space 
\begin{equation}
W=W_{anisotropic}({\bf C}_S, {\bf b}_0, {\bf p})=W_{isotropic}({\bf C}_S, {\bf b}_0, {\bf P}_6, {\bf p}). 
\end{equation}
The structure tensor for $\mathcal D_{6h}$ is denoted by ${\bf P}_6$ and defined by (\cite{Zheng1993})
\begin{equation}
{\bf P}_6=Re({\bf i}+i {\bf j})^6,
\end{equation}
or equivalently as 
\begin{equation}
{\bf P}_6={\bf U} \otimes {\bf U} \otimes {\bf U}-({\bf L} \otimes {\bf U}  \otimes {\bf L}+{\bf L} \otimes {\bf L} \otimes {\bf U}),
\end{equation}
where ${\bf U}={\bf a}_1 \otimes {\bf a}_1- {\bf a}_2 \otimes {\bf a}_2$, ${\bf L}={\bf a}_1 \otimes {\bf a}_2 - {\bf a}_2 \otimes {\bf a}_1$, ${{\bf a}_1, {\bf a}_2}$, $\{ {\bf a}_1, {\bf a}_2  \}$ an orthonormal basis vector. It can also be written as  
\begin{equation}
{\bf P}_6=Re[e^{i 6 \theta} ({\bf c}_1+i {\bf c}_2)^6],
\end{equation} 
where ${\bf c}_1=cos(\theta){\bf a}_1+sin(\theta) {\bf a}_2$, ${\bf c}_2=-sin(\theta){\bf a}_1+cos(\theta) {\bf a}_2$. This tensor is an irreducible tensor since $\mathcal D_{6h}$ is compact. In two dimensions it has only two independent components (\cite{Zheng1994})
\begin{equation}
{\bf P}_{111111}=\textrm{cos}(6 \theta), \ \ {\bf P}_{211111}=\textrm{sin}(6 \theta).
\end{equation}
These two components introduce the anisotropy and for graphene, they model the zig-zag and armchair direction. Since $\theta=\frac{2 \pi}{6}$ we have $P_{111111}=cos(2 \pi)=1, P_{211111}=sin(2 \pi)=0 $. 

Thus, we take an isotropic function at the expense of using the structure tensor as an additional argument. The complete and irreducible representation of such a scalar function under the group $\mathcal D_{6h}$ consists of the thirteen invariants (\cite{Zheng1994,Zheng1993})
\begin{eqnarray}
&&I_1=\textrm{tr}{\bf C}_S, \ I_2=\textrm{det}{\bf C}_S, \ I_3=\textrm{tr}(\Pi^{{\bf C}_S}_6 {\bf C}_S ), \ I_4=\textrm{tr}({\bf C}_S {\bf b}_0), \nonumber\\
&&I_5=\textrm{tr}(\Pi^{{\bf b}_0}_6 {\bf b}_0), \  I_6=\textrm{tr}{\bf b}_0, \ I_7=\textrm{det}{\bf b}_0, \ I_8={\bf p} \cdot {\bf C}_S {\bf p}, \nonumber\\
&&I_9={\bf p} \cdot {\bf b}_0 {\bf p}, \  I_{10}={\bf p} \cdot \pi^{{\bf p}}_6, \ I_{11}={\bf p} \cdot {\bf p}, \ I_{12}=\textrm{tr}(\Pi^{\bf p}_6 {\bf C}_S), \ I_{13}=\textrm{tr}(\Pi^{\bf p}_6 {\bf b}_0).
\end{eqnarray}

So, in general for such a model of graphene we have the following expression for the energy 
\begin{equation}
W=\tilde{W}(I_i), \ i=1,2,...,13. 
\end{equation}
The term $\Pi^{\bf A}_6$ for a symmetric tensor of second order $\bf A$ is defined as (\cite{Zheng1994}, with indices ranging from 1 to 2)
\begin{equation}
\Pi^{\bf A}_6=P_{ijklmn} A_{kl} A_{mn} {\bf c}_i \otimes {\bf c}_j,
\end{equation}
and renders a second order tensor. For the definition of the basis $\{{\bf c}_1, {\bf c}_2 \}$ see \cite{Zheng1994,Zheng1993,Sfyris-GaliotisMMS}. The term $\pi^{\bf z}_6$ with respect to the vector $\bf z$ is defined as 
\begin{equation}
\pi^{\bf z}_6=P_{ijklmn} z_j z_k z_l z_m z_n {\bf c}_i,
\end{equation}
while for $\Pi^{\bf z}_6$ we have
\begin{equation}
\Pi^{\bf z}_6=P_{ijklmn} z_k z_l z_m z_n {\bf c}_i \otimes {\bf c}_j.
\end{equation}

The material parameters related to $I_6, I_7$ describe pure bending effects since det${\bf b}_0$, tr${\bf b}_0$ are the mean and the Gaussian curvature of the surface, respectively. The term related with $I_5$ describes the effect of the armchair and the zigzag direction of graphene at bending. The parameters related with $I_1, I_2$ are related to pure stretching, while those related with the term $I_3$ describe the effect of anisotropy (zigzag, armchair directions) to stretching. The parameter related to $I_4$ describes coupling between bending and stretching responses. The terms $I_8, I_9$ describe the effect of the in plane and the out of plane deformations, respectively, on the shift vector. The term $I_{10}$ describes the way anisotropy affects the shift vector, while $I_{11}$ describes changes related with the shift vector solely. Terms $I_{12}, I_{13}$ are related to coupling of anisotropy with the shift vector for the in plane and the out of plane deformations, respectively. The zigzag and armchair directions of a graphene sheet are depicted in Figures 3, 4. 
\begin{figure}[!htb]
\centering
\includegraphics{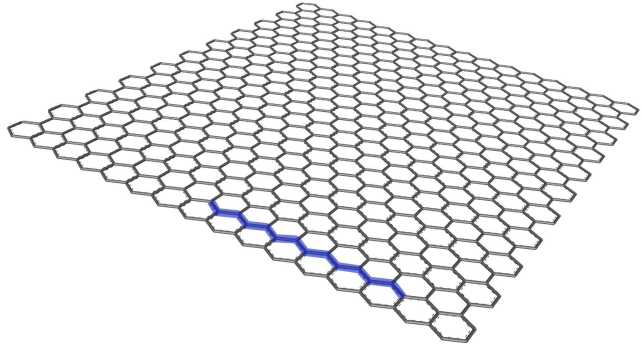}
\caption{The armchair direction of graphene is introduced into the mathematical framework through the tensors $\Pi^{{\bf C}_S}_6, \Pi^{{\bf b}_0}_6, \pi^{\bf p}_6, \Pi^{\bf p}_6$.}
\label{fig:digraph}
\end{figure}
\begin{figure}[!htb]
\centering
\includegraphics{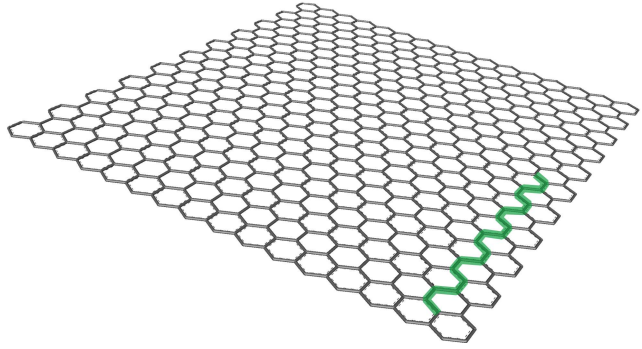}
\caption{The zigzag direction of graphene is introduced into the mathematical framework through the tensors $\Pi^{{\bf C}_S}_6, \Pi^{{\bf b}_0}_6, \pi^{\bf p}_6, \Pi^{\bf p}_6$.}
\label{fig:digraph}
\end{figure}

For evaluating the surface stress tensor and the surface couple stress tensor one has to determine the derivatives of the energy with respect to the Cauchy-Green tensor and the curvature tensor: 
\begin{equation}
{\bf S}_S=\frac{\partial \tilde{W}}{\partial {\bf C}_S}, \ \ {\bf m}_S=\frac{\partial \tilde{W}}{\partial {\bf b}_0}.
\end{equation}
Also the term $\frac{\partial \tilde{W}}{\partial {\bf p}}$ is important, since it is present to the field equations ruling the auxiliary variables. Using the expressions of eq. (22) in eq. (27), after some calculations we obtain 
\begin{eqnarray}
{\bf S}_S&=&\frac{\partial \tilde{W}}{\partial I_1} {\bf G} + \frac{\partial \tilde{W}}{\partial I_2} [ \textrm{tr}({\bf C}_S) {\bf 1} -{\bf C}_S ]+3 \frac{\partial \tilde{W}}{\partial I_3} {\bf P}_6 : ({\bf C}_S \otimes  {\bf C}_S)+ \frac{\partial \tilde{W}}{\partial I_4} {\bf b}_0 \nonumber\\ 
&&+\frac{\partial \tilde{W}}{\partial I_8} {\bf p} \otimes {\bf p}+\frac{\partial \tilde{W}}{\partial I_{12}}\Pi^{\bf p}_6, \\
{\bf m}_S&=&\frac{\partial \tilde{W}}{\partial I_6} {\bf G} + \frac{\partial \tilde{W}}{\partial I_7} [ \textrm{tr}({\bf b}_0) {\bf 1} -{\bf b}_0 ]+3 \frac{\partial \tilde{W}}{\partial I_5} {\bf P}_6 : ({\bf b}_0 \otimes  {\bf b}_0)+ \frac{\partial \tilde{W}}{\partial I_4} {\bf C}_S \nonumber\\
&& +\frac{\partial \tilde{W}}{\partial I_9} {\bf p} \otimes {\bf p}+\frac{\partial \tilde{W}}{\partial I_{13}} \Pi^{\bf p}_6, \\
\frac{\partial W}{\partial {\bf p}}&=&2\frac{\partial \tilde{W}}{\partial I_8} {\bf C}_S {\bf p}+2\frac{\partial \tilde{W}}{\partial I_9} {\bf b}_0 {\bf p}+6 \frac{\partial \tilde{W}}{\partial I_{10}} {\bf P}_6 \bullet ({\bf p} \otimes {\bf p} \otimes {\bf p} \otimes {\bf p} \otimes {\bf p}) + \frac{\partial \tilde{W}}{\partial I_{11}} {\bf p} \nonumber\\
&&+4 \frac{\partial \tilde{W}}{\partial I_{12}} [{\bf P}_6 : ( {\bf p} \otimes {\bf p} \otimes {\bf p})]: {\bf C}_S+4 \frac{\partial \tilde{W}}{\partial I_{13}} [{\bf P}_6 : ( {\bf p} \otimes {\bf p} \otimes {\bf p})]: {\bf b}_0.
\end{eqnarray}
The referential metric tensor is denoted by $\bf G$. By making the simplest possible assumption that $\tilde{W}$ is linear with respect to the invariants $I_i, i=1,2,3,...,13$ we take 
\begin{eqnarray}
{\bf S}_S&=&\alpha{\bf G} + \beta [ \textrm{tr}({\bf C}_S) {\bf 1} -{\bf C}_S ]+3 \gamma {\bf P}_6 : ({\bf C}_S \otimes  {\bf C}_S)+ \delta {\bf b}_0+\theta  {\bf p} \otimes {\bf p}+\rho \Pi^p_6, \\
{\bf m}_S&=&\epsilon {\bf G} + \zeta [ \textrm{tr}({\bf b}_0) {\bf 1} -{\bf b}_0 ]+3 \eta {\bf P}_6 : ({\bf b}_0 \otimes  {\bf b}_0)+ \delta {\bf C}_S+\iota {\bf p} \otimes {\bf p}+\tau \Pi^p_6, \\
\frac{\partial W}{\partial {\bf p}}&=&\theta {\bf C}_S {\bf p}+\iota {\bf b}_0 {\bf p}+6 \lambda {\bf P}_6 \bullet( {\bf p} \otimes {\bf p} \otimes {\bf p} \otimes {\bf p} \otimes {\bf p})+\xi {\bf p} \nonumber\\
&& +4 \rho  [{\bf P}_6 : ( {\bf p} \otimes {\bf p} \otimes {\bf p})]: {\bf C}_S +4 \tau [{\bf P}_6 : ( {\bf p} \otimes {\bf p} \otimes {\bf p})]: {\bf b}_0.
\end{eqnarray}
The Greek letters $\alpha, \beta, \gamma, \delta, \theta, \rho, \epsilon, \zeta, \eta, \iota, \tau, \lambda, \xi$ symbolize the thirteen different material parameters that can be determined by experiments. 

Using index notation, with indices ranging from 1 to 2, the above equations read
\begin{eqnarray}
S_{S_{AB}}&=&\alpha {\bf G}_{AB}+\beta [\textrm{tr} ({\bf C}_S) \delta_{AB}-C_{S_{AB}}]+3 \gamma P_{ABCDEF} C_{S_{EF}} C_{S_{CD}} +\delta b_{0_{AB}} \nonumber\\
&& +\theta p_A p_B +\rho P_{ABCDEF} p_C p_D p_E p_F, \\
m_{S_{AB}}&=&\epsilon {\bf G}_{AB}+\zeta [\textrm{tr} ({\bf b}_0) \delta_{AB}-b_{0_{AB}}]+3 \eta P_{ABCDEF} b_{0_{EF}} b_{0_{CD}} +\delta C_{S_{AB}} \nonumber\\
&& +\iota p_A p_B +\tau P_{ABCDEF} p_C p_D p_E p_F, \\
\frac{\partial W}{\partial p_A}&=& \theta C_{S_{AB}} p_A +\iota b_{0_{AB}} p_B+6 \lambda P_{ABCDEF} p_B p_C p_D p_E p_F+\xi p_A \nonumber\\
&&+4 \rho P_{ABCDEF} p_D p_E p_F C_{S_{BC}}+4 \tau P_{ABCDEF} p_D p_E p_F b_{0_{BC}}. 
\end{eqnarray}

The elasticities of this model are given by the following fourth order tensors
\begin{equation}
\mathcal A=\frac{\partial^2 W}{\partial {\bf C}_S^2}, \ \ \mathcal B=\frac{\partial^2 W}{\partial {\bf b}_0^2}, \ \ \mathcal C=\frac{\partial^2 W}{\partial {\bf C}_S \partial {\bf b}_0}.
\end{equation}
Quantities of the first term are related to the in-plane motion, the second term related to the out-of-plane motion while the third term is related to the coupling between in-plane and out-of-plane motions.

The field equations for such a problem are the momentum equation, the moment of momentum equation as well as the equations ruling the shift vector. For the momentum equation we have (\cite{Chhapadiaetal2011,Sfyris-GaliotisMMS}) when body forces and inertia are absent
\begin{equation}
\boldsymbol \sigma^{\textrm{bulk}} \cdot {\bf n}+\nabla_S {\bf T}_S=0,
\end{equation}
where ${\bf T}_S$ is the surface first Piola-Kirchhoff stress tensor defined by (\cite{Steigmann-Ogden1999})
\begin{equation}
{\bf T}_S=\frac{\partial \bar{W}}{\partial {\bf F}_S}
\end{equation}
when $W=\bar{W}({\bf F}_S, {\bf b}_0, {\bf p})$, while $\boldsymbol \sigma^{\textrm{bulk}}$ is the Cauchy stress tensor for the bulk material surrounding the surface. The surface divergence $\nabla_S()$ for a quantity is defined as
\begin{equation}
\nabla_S()=\nabla()-{\bf n} ({\bf n} \cdot \nabla()),
\end{equation}
$\bf n$ being the outward unit normal to the surface at hand. Here since we speak about a free standing surface, there is no bulk material surrounding the graphene, so the bulk stress tensor should be set equal to zero, $\boldsymbol \sigma^{\textrm{bulk}}={\bf 0}$. In this case the momentum equation reads
\begin{equation}
\nabla_S {\bf T}_S= {\bf 0}.
\end{equation}

The surface first Piola-Kirchhoff stress tensor is related to the second Piola-Kirchhoff surface stress tensor, ${\bf S}_S$, according to the formula
\begin{equation}
{\bf S}_S={\bf F}_S^{-1} \cdot  {\bf T}_S.
\end{equation} 

The moment of momentum balance, in the absence of body couples and inertia reads (\cite{Chhapadiaetal2011})
\begin{equation}
{\bf x} \times ({\bf n} \cdot {\boldsymbol \sigma}^{\textrm{bulk}})+\nabla_S ({\bf F}_S \cdot {\bf m}_S)-\nabla_S ({\bf F}_S \cdot  {\bf S}_S \times {\bf x})={\bf 0}.
\end{equation}
When the graphene monolayer is free standing we set $\boldsymbol \sigma^{\textrm{bulk}}={\bf 0}$ to obtain
\begin{equation}
\nabla_S ({\bf F}_S \cdot {\bf m}_S)-\nabla_S ({\bf F}_S \cdot  {\bf S}_S \times {\bf x})={\bf 0}. 
\end{equation}
The symbol $\times$ in eq. (43) denoted the cross product of the three dimensional space.

For the shift vector the field equation reads (\cite{Pitteri-Zanzotto2003,E-Ming2007})
\begin{equation}
\frac{\partial W}{\partial {\bf p}}={\bf 0}.
\end{equation}
Form the physical point of view, the momentum equation is the force balance for the surface, while the moment of momentum renders the couple balance for the surface. The shift vector adjusts according to eq. (45) in order equilibrium to be reached (\cite{Pitteri-Zanzotto2003}). 

\section{Field equations in terms of the kinematic quantities}

The following three quantities: $\{ {\bf x}, {\bf b}, {\bf p}  \}$, i.e., the position vector, the curvature tensor and the shift vector constitute the solution of the free standing monolayer graphene sheet problem. Their calculation comes from solving a system of equations: the momentum equations, eq. (41), the moment of momentum equations, eq. (44) and the equations ruling the shift vector, eq. (45). For the first of them we have 
\begin{eqnarray}
&&\{ x_{1,1}[\alpha G_{11}+\beta (x^2_{1,2}+x^2_{2,2}+x^2_{3,2})+3\gamma (x^2_{1,1}+x^2_{2,1}+x^2_{3,1})^2+\delta b_{11}+\theta p_1^2+\rho p_1^4] \}_{,1}+ \nonumber\\
&&\{ x_{1,1}[\alpha G_{12}-\beta (x_{1,1} x_{1,2}+x_{2,1} x_{2,2}+x_{3,1} x_{3,2})+\delta b_{12} +\theta p_1 p_2] \}_{,2}+  \nonumber\\
&&\{ x_{1,2}[\alpha G_{12}-\beta (x_{1,1} x_{1,2}+x_{2,1} x_{2,2}+x_{3,1} x_{3,2}) +\delta b_{12} +\theta p_1 p_2 ] \}_{,1}+ \nonumber\\
&&\{ x_{1,2}[\alpha G_{22}+\beta(x^2_{1,1}+x^2_{2,1}+x^2_{3,1})+\delta b_{22}+\theta p^2_2] \}_{,2}=0.
\end{eqnarray}
For the second we obtain  
\begin{eqnarray}
&&\{ x_{2,1}[\alpha G_{11}+\beta (x^2_{1,2}+x^2_{2,2}+x^2_{3,2})+3\gamma (x^2_{1,1}+x^2_{2,1}+x^2_{3,1})^2+\delta b_{11}+\theta p_1^2+\rho p_1^4] \}_{,1}+ \nonumber\\
&&\{ x_{2,1}[\alpha G_{12}-\beta (x_{1,1} x_{1,2}+x_{2,1} x_{2,2}+x_{3,1} x_{3,2})+\delta b_{12} +\theta p_1 p_2] \}_{,2}+  \nonumber\\
&&\{ x_{2,2}[\alpha G_{12}-\beta (x_{1,1} x_{1,2}+x_{2,1} x_{2,2}+x_{3,1} x_{3,2}) +\delta b_{12} +\theta p_1 p_2 ] \}_{,1}+ \nonumber\\
&&\{ x_{2,2}[\alpha G_{22}+\beta(x^2_{1,1}+x^2_{2,1}+x^2_{3,1})+\delta b_{22}+\theta p^2_2] \}_{,2}=0.
\end{eqnarray}

The first of the equations ruling the auxiliary variables, eq. (45), read
\begin{eqnarray}
&&\theta (x^2_{1,1}+x^2_{2,1}+x^2_{3,1})p_1+\theta (x_{1,1} x_{1,2}+x_{2,1} x_{2,2}+x_{3,1} x_{3,2})p_2+\iota b_{11} p_1 +\iota b_{12} p_2 \nonumber\\
&&+6 \lambda p_1^5 +\xi p_1+4 \rho p^3_1(x^2_{1,1}+x^2_{2,1}+x^2_{3,1})+4 \tau p^3_1 b_{11}=0,
\end{eqnarray}
while the second renders 
\begin{eqnarray}
&&\theta (x_{1,1} x_{1,2} +x_{2,1} x_{2,2}+x_{3,1} x_{3,2})p_1+\theta (x^2_{1,2}+x^2_{2,2}+x^2_{3,2})p_2+\iota b_{21} p_1+\iota b_{22} p_2+\xi p_2=0.
\end{eqnarray}

The equations of moment of momentum are found by using eq. (34, 35, 42) on eq. (44). For the first equation of moment of momentum we then have
\begin{eqnarray}
&& \{ x_{1,1} [\epsilon G_{11} +\zeta b_{22}+3 \eta (x^2_{1,1}+x^2_{2,1}+x^2_{3,1})^2+\delta (x^2_{1,1}+x^2_{2,1}+x^2_{3,1})+\iota p_1^2+\tau p_1^4  ]    \}_{,1} \nonumber\\
&&+\{ x_{1,2} [\epsilon G_{21}-\zeta b_{12}+\delta (x_{1,1} x_{1,2}+x_{2,1}x_{2,2}+x_{3,1}x_{3,2})+\iota p_1 p_2  ]  \}_{,1} \nonumber\\
&&+\{ x_{1,1} [\epsilon G_{12} -\zeta b_{12} +\delta (x_{1,1} x_{1,2}+x_{2,1}x_{2,2}+x_{3,1}x_{3,2})+\iota p_1 p_2  ]   \}_{,2} \nonumber\\
&&+\{ x_{1,2} [\epsilon G_{22}+\zeta b_{11}+\delta(x^2_{1,2}+x^2_{2,2}+x^2_{3,2})+\iota p_2^2    ]   \}_{,2}  \nonumber\\
&&-x_{1,1} \{   x_{1,1}[\alpha G_{12} -\beta (x_{1,1} x_{1,2}+x_{2,1}x_{2,2}+x_{3,1}x_{3,2})+\delta b_{12} +\theta p_1 p_2   ] \nonumber\\
&& \ \ \ \ \ \ \ \ \ \ +x_{1,2} [ \alpha G_{22} +\beta (x^2_{1,1}+x^2_{2,1}+x^2_{3,1}) +\delta b_{22}+\theta p_2^2   ] \} \nonumber\\
&& -x_{1,2} \{   x_{1,1}[\alpha G_{12} -\beta (x_{1,1} x_{1,2}+x_{2,1}x_{2,2}+x_{3,1}x_{3,2})+\delta b_{12} +\theta p_1 p_2   ] \nonumber\\
&& \ \ \ \ \ \ \ \ \ \ +x_{1,2} [ \alpha G_{22} +\beta (x^2_{1,1}+x^2_{2,1}+x^2_{3,1}) +\delta b_{22}+\theta p_2^2   ] \} \nonumber\\   
&&+x_{2,1} \{   x_{1,1}[ \alpha G_{11} +\beta (x^2_{1,2}+x^2_{2,2}+x^2_{3,2})+3 \gamma (x^2_{1,1}+x^2_{2,1}+x^2_{3,1})^2 +\delta b_{11}+\theta p_1^2+\rho p_1^4  ] \nonumber\\
&&  \ \ \ \ \ \ \ \ \  +x_{1,2} [\alpha G_{21} -\beta (x_{1,1} x_{1,2}+x_{2,1}x_{2,2}+x_{3,1}x_{3,2}) +\delta b_{12} +\theta p_1 p_2   ]   \}\nonumber\\
&&+x_{2,2} \{   x_{1,1}[ \alpha G_{11} +\beta (x^2_{1,2}+x^2_{2,2}+x^2_{3,2})+3 \gamma (x^2_{1,1}+x^2_{2,1}+x^2_{3,1})^2 +\delta b_{11}+\theta p_1^2+\rho p_1^4  ] \nonumber\\
&&  \ \ \ \ \ \ \ \ \  +x_{1,2} [\alpha G_{21} -\beta (x_{1,1} x_{1,2}+x_{2,1}x_{2,2}+x_{3,1}x_{3,2}) +\delta b_{12} +\theta p_1 p_2   ]   \}\nonumber\\                     
&&-x_1  \{   x_{1,1}   [ \alpha G_{12}-\beta  (x_{1,1} x_{1,2}+x_{2,1}x_{2,2}+x_{3,1}x_{3,2})+\delta b_{12} +\theta p_1 p_2    ] \nonumber\\
&&\ \ \ \ \ \ \ \ \ \  +x_{1,2} [ \alpha G_{22} +\beta (x^2_{1,1}+x^2_{2,1}+x^2_{3,1})+\delta b_{22}+\theta p_2^2  ]  \}_{,1} \nonumber\\
&&-x_1  \{   x_{1,1}   [\alpha G_{12} -\beta  (x_{1,1} x_{1,2}+x_{2,1}x_{2,2}+x_{3,1}x_{3,2})+\delta b_{12} +\theta p_1 p_2    ] \nonumber\\
&&\ \ \ \ \ \ \ \ \ \  +x_{1,2} [ \alpha G_{22} +\beta (x^2_{1,1}+x^2_{2,1}+x^2_{3,1})+\delta b_{22}+\theta p_2^2  ]  \}_{,2} \nonumber\\
&&+x_2 \{    x_{1,1}  [ \alpha G_{11} +\beta (x^2_{1,2}+x^2_{2,2}+x^2_{3,2}) + 3 \gamma (x^2_{1,1}+x^2_{2,1}+x^2_{3,1})^2 + \delta b_{11}  +\theta p_1^2 +\rho p_1^4   ]   \nonumber\\
&&\ \ \ \ \ \ \ \ \ \ +x_{1,2} [\alpha G_{21}-\beta (x_{1,1} x_{1,2}+x_{2,1}x_{2,2}+x_{3,1}x_{3,2}) +\delta b_{12}+\theta p_1 p_2   ]  \}_{,1} \nonumber\\
&&+x_2 \{    x_{1,1}  [ \alpha G_{11} +\beta (x^2_{1,2}+x^2_{2,2}+x^2_{3,2}) + 3 \gamma (x^2_{1,1}+x^2_{2,1}+x^2_{3,1})^2 + \delta b_{11}  +\theta p_1^2 +\rho p_1^4   ]   \nonumber\\
&&\ \ \ \ \ \ \ \ \ \  +x_{1,2} [\alpha G_{21} -\beta (x_{1,1} x_{1,2}+x_{2,1}x_{2,2}+x_{3,1}x_{3,2}) +\delta b_{12}+\theta p_1 p_2   ]  \}_{,2}=0.
\end{eqnarray}
The second equation of the moment of momentum becomes
\begin{eqnarray}
&& \{   x_{2,1}[\epsilon G_{11} +\zeta b_{22}+3 \eta (x^2_{1,1}+x^2_{2,1}+x^2_{3,1})^2 +\delta (x^2_{1,1}+x^2_{2,1}+x^2_{3,1})+\iota p_1^2+\tau p_1^4  ]    \}_{,1} \nonumber\\
&&+\{   x_{2,2} [\epsilon G_{21} -\zeta b_{12} +\delta (x_{1,1} x_{1,2}+x_{2,1}x_{2,2}+x_{3,1}x_{3,2}) + \iota p_1 p_2   ]  \}_{,1} \nonumber\\
&&+\{   x_{2,1} [\epsilon G_{12} -\zeta b_{12} +\delta (x_{1,1} x_{1,2}+x_{2,1}x_{2,2}+x_{3,1}x_{3,2}) + \iota p_1 p_2   ]  \}_{,2} \nonumber\\
&&+\{   x_{2,2} [\epsilon G_{22} +\zeta b_{11}  +\delta (x^2_{1,2}+x^2_{2,2}+x^2_{3,2})+\iota p_2^2    ]    \}_{,2} \nonumber\\
&&-x_{1,1}  \{   x_{2,1}  [\alpha G_{12} -\beta (x_{1,1} x_{1,2}+x_{2,1}x_{2,2}+x_{3,1}x_{3,2}) +\delta b_{12} +\theta p_1 p_2    ] \nonumber\\
&&\ \ \ \ \ \ \ \ \ \  +x_{2,2} [ \alpha G_{22} +\beta (x^2_{1,1}+x^2_{2,1}+x^2_{3,1})+\delta b_{22}+\theta p_2^2   ]  \} \nonumber\\
&&-x_{1,2}  \{   x_{2,1}  [\alpha G_{12} -\beta (x_{1,1} x_{1,2}+x_{2,1}x_{2,2}+x_{3,1}x_{3,2}) +\delta b_{12} +\theta p_1 p_2    ] \nonumber\\
&&\ \ \ \ \ \ \ \ \ \  +x_{2,2} [ \alpha G_{22} +\beta (x^2_{1,1}+x^2_{2,1}+x^2_{3,1})+\delta b_{22}+\theta p_2^2   ]  \} \nonumber\\
&&+x_{2,1} \{  x_{2,1} [ \alpha G_{11} +\beta (x^2_{1,2}+x^2_{2,2}+x^2_{3,2})  +3 \gamma (x^2_{1,1}+x^2_{2,1}+x^2_{3,1})^2 +\delta b_{11}+\theta p_1^2+\rho p_1^4  ] \nonumber\\
&& \ \ \ \ \ \ \ \ \ \ +x_{2,2} [\alpha G_{21} -\beta (x_{1,1} x_{1,2}+x_{2,1}x_{2,2}+x_{3,1}x_{3,2}) +\delta b_{12}+\theta p_1 p_2  ]  \} \nonumber\\
&&+x_{2,2} \{  x_{2,1} [ \alpha G_{11}+\beta (x^2_{1,2}+x^2_{2,2}+x^2_{3,2})  +3 \gamma (x^2_{1,1}+x^2_{2,1}+x^2_{3,1})^2 +\delta b_{11}+\theta p_1^2+\rho p_1^4  ] \nonumber\\
&& \ \ \ \ \ \ \ \ \ \ +x_{2,2} [\alpha G_{21} -\beta (x_{1,1} x_{1,2}+x_{2,1}x_{2,2}+x_{3,1}x_{3,2}) +\delta b_{12}+\theta p_1 p_2  ]  \} \nonumber\\
&&-x_1\{  x_{2,1} [ \alpha G_{12}-\beta (x_{1,1} x_{1,2}+x_{2,1}x_{2,2}+x_{3,1}x_{3,2}) +\delta b_{12} +\theta p_1 p_2   ] \nonumber\\
&&\ \ \ \ \ \ \ \ \ \ \ +x_{2,2} [ \alpha G_{22} +\beta (x^2_{1,1}+x^2_{2,1}+x^2_{3,1})+\delta b_{22}+\theta p_2^2  ] \}_{,1} \nonumber\\ 
&&-x_1\{  x_{2,1} [\alpha G_{12} -\beta (x_{1,1} x_{1,2}+x_{2,1}x_{2,2}+x_{3,1}x_{3,2}) +\delta b_{12} +\theta p_1 p_2   ] \nonumber\\
&&\ \ \ \ \ \ \ \ \ \ \ +x_{2,2} [ \alpha G_{22} +\beta (x^2_{1,1}+x^2_{2,1}+x^2_{3,1})+\delta b_{22}+\theta p_2^2  ] \}_{,2} \nonumber\\
&&+x_2  \{  x_{2,1} [ \alpha G_{11} +\beta (x^2_{1,2}+x^2_{2,2}+x^2_{3,2}) +3 \gamma (x^2_{1,1}+x^2_{2,1}+x^2_{3,1})^2 +\delta b_{11}+\theta p_1^2+\rho p_1^4     ] \nonumber\\
&&\ \ \ \ \ \ \ \ \ \ \ +x_{2,2} [\alpha G_{21} -\beta (x_{1,1} x_{1,2}+x_{2,1}x_{2,2}+x_{3,1}x_{3,2}) +\delta b_{12}+\theta p_1 p_2   ]  \}_{,1} \nonumber\\
&&+x_2  \{  x_{2,1} [ \alpha G_{11} +\beta (x^2_{1,2}+x^2_{2,2}+x^2_{3,2}) +3 \gamma (x^2_{1,1}+x^2_{2,1}+x^2_{3,1})^2 +\delta b_{11}+\theta p_1^2+\rho p_1^4     ] \nonumber\\
&&\ \ \ \ \ \ \ \ \ \ \ +x_{2,2} [\alpha G_{21} -\beta (x_{1,1} x_{1,2}+x_{2,1}x_{2,2}+x_{3,1}x_{3,2}) +\delta b_{12}+\theta p_1 p_2   ]  \}_{,2}=0. \nonumber\\
\end{eqnarray}

\section{In-plane loading of a monolayer graphene}

As a first approach, we examine a simplified model that is capable of describing bending as well as the effect of anisotropy to both bending and in-plane deformations. At the constitutive law we assume dependence of the energy on the invariants $I_1, I_3, I_5, I_6, I_8, I_9, I_{10}, I_{11}$; this means that the elastic constants $\beta, \delta, \rho, \zeta, \tau$ are equal to zero. Also, we assume that at the reference configuration the body is a rectangle with length $L_1$ and width $L_2$: $-L_1 \leq \Theta^1 \leq L_1, -L_2 \leq \Theta^2 \leq L_2$. For the components of the referential metric tensor, $\bf G$, we then have $G_{11}=G_{22}=1, G_{12}=G_{21}=0$. The further assumption that there are no out of plane motions, translates to $x_3=ct$, where $ct$ is a constant, for the position vector ${\bf x}={\bf x}({\Theta^1, \Theta^2})$. Thus, all derivatives of $x_3$ appearing to the field equations are zero and in addition to that, the moment of momentum equations are satisfied trivially. Essentially, with these set of assumptions, there is no longer dependence of the energy on the curvature tensor. 

The two equations of momentum for this model reads
\begin{eqnarray}
&&\{ x_{1,1}[\alpha +3\gamma (x^2_{1,1}+x^2_{2,1})^2+\theta p_1^2] \}_{,1}+\{ x_{1,1}[\theta p_1 p_2] \}_{,2}+  \nonumber\\
&&\{ x_{1,2}[\theta p_1 p_2 ] \}_{,1}+ \{ x_{1,2}[\alpha +\theta p^2_2] \}_{,2}=0,
\end{eqnarray}
\begin{eqnarray}
&&\{ x_{2,1}[\alpha +3\gamma (x^2_{1,1}+x^2_{2,1})^2+\theta p_1^2] \}_{,1}+\{ x_{2,1}[\theta p_1 p_2] \}_{,2}+  \nonumber\\
&&\{ x_{2,2}[\theta p_1 p_2 ] \}_{,1}+ \{ x_{2,2}[\alpha +\theta p^2_2] \}_{,2}=0.
\end{eqnarray}
For the equations ruling the auxiliary variables we obtain 
\begin{eqnarray}
&&\theta (x^2_{1,1}+x^2_{2,1}) p_1 + \theta (x_{1,1} x_{1,2}+x_{2,1} x_{2,2}) p_2 +6 \lambda p_1^5 +\xi p_1=0
\end{eqnarray}
\begin{eqnarray}
&& \theta (x_{1,1} x_{1,2} +x_{2,1} x_{2,2}) p_1  + \theta (x^2_{1,2}+x^2_{2,2}) p_2=0.
\end{eqnarray}
Using this approach we come to a system of equations that is possible to be solved for specific types of mechanical loading. 

\subsection{Tension/Compression}

The first type of mechanical loading that we apply to the graphene sheet is the one dimensional tension/compression. There are two possible ways for tensile/compressive experiments to be conducted since the material is essentially a two dimensional surface. The first way is to have the following form for the in-plane deformation
\begin{eqnarray}
&& x_1=\varepsilon \Theta^1 \nonumber\\
&& x_2=\Theta^2.
\end{eqnarray} 
This is a homogeneous deformation for every material point in the direction of $x_1$, as is seen schematically in Figure 5.  
\begin{figure}[!htb]
\centering
\includegraphics{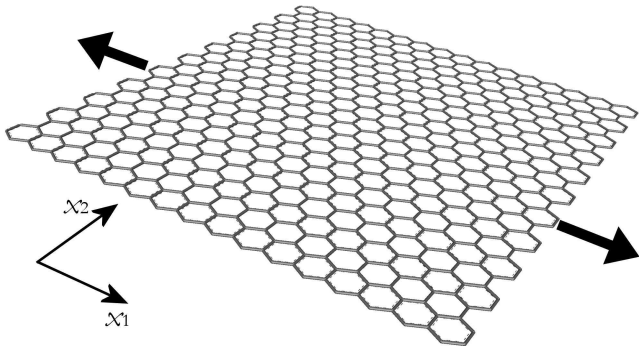}
\caption{In-plane tension along the $x_1$ direction.}
\label{fig:digraph}
\end{figure}
The body experiences tension if the mechanical loading constant is positive, i.e. $\varepsilon > 0$, while compression appears if the constant is negative, i.e. $\varepsilon < 0$. The derivatives of the position vector then reads
\begin{equation}
x_{1,1}=\varepsilon, \ \ x_{1,2}=0, \ \ x_{2,2}=1, \ \ x_{2,1}=0.
\end{equation} 
Substitution of these forms to eqs. (52-55) gives the following system of equations that should be satisfied in order that the deformation of eq. (56) to be a solution
\begin{eqnarray}
&& 2 \varepsilon \theta p_1 p_{1,1} + \varepsilon \theta p_{1,2} p_2+ \varepsilon \theta p_1 p_{2,2}=0, \\
&& \theta p_{1,1} p_2 +\theta p_1 p_{2,1} +2 \theta p_2 p_{2,2}=0, \\
&& \varepsilon^2 \theta p_1+ 6\lambda p_1^5+\xi p_1=0, \\
&& \theta p_2+\xi p_2=0.
\end{eqnarray} 
The unknown quantities of this system are the components of the shift vector that, in general, are functions of $\Theta^1$ and $\Theta^2$; for brevity we write $p_1=p_1(\Theta^1, \Theta^2)$ and $p_2=p_2(\Theta^1, \Theta^2)$.

Eq. (60) can be written in the form $6 \lambda p_1^5+(\theta \epsilon^2+\xi)p_1=0$ which has the following five solutions for $p_1$
\begin{eqnarray}
&& p_1=0, \ \ p_1=-\frac{(-1)^{1/4} (\theta \varepsilon^2+\xi)^{1/4}}{(6 \lambda)^{1/4}}, \ \  p_1=\frac{(-1)^{1/4} (\theta \varepsilon^2+\xi)^{1/4}}{(6 \lambda)^{1/4}}, \nonumber\\
&&  p_1=-\frac{(-1)^{3/4} (\theta \varepsilon^2+\xi)^{1/4}}{(6 \lambda)^{1/4}}, \ \  p_1=\frac{(-1)^{3/4} (\theta \varepsilon^2+\xi)^{1/4}}{(6 \lambda)^{1/4}}.
\end{eqnarray}
Eqs. (62)$_{2,3}$ can be written as
\begin{equation}
p_1=\pm (-1)^{1/4} \left( \frac{(\theta \varepsilon^2+\xi)}{6 \lambda} \right)^{1/4}= \pm \sqrt{i} \sqrt{ \left( \frac{(\theta \varepsilon^2+\xi)}{6 \lambda} \right)^{1/2}}.
\end{equation}
When $(\theta \varepsilon^2+\xi) \lambda > 0 $ the above two solutions render $p_1$ a complex number, so a physically not-acceptable solution. In the case when $(\theta \varepsilon^2+\xi) \lambda < 0 $ a similar situation holds. To see this we set $\frac{(\theta \varepsilon^2+\xi)}{6 \lambda}=-f, f>0$. Then we have $\left( \frac{(\theta \varepsilon^2+\xi)}{6 \lambda} \right)^{1/2}=(-f)^{1/2}=(-1)^{1/2}f^{1/2}=i f^{1/2}$. So, collectively we obtain $p_1=\pm \sqrt{i} \sqrt{(-f)^{1/2}}=\pm \sqrt{i} \sqrt{i} \sqrt{f^{1/2}}=\pm i \sqrt{f^{1/2}}$, which means that $p_1$ is a complex number. 

The last two solutions of eq. (62)$_{4,5}$ render $p_1$ a real number when $(\theta \varepsilon^2+\xi) \lambda < 0$. To see this we set again $\frac{(\theta \varepsilon^2+\xi)}{6 \lambda}=-f, f>0$ and we get
\begin{equation}
p_1=\pm \frac{(-1)^{3/4} (\theta \varepsilon^2+\xi)^{1/4}}{(6 \lambda)^{1/4}}=\pm \sqrt{i^3}  \left( \frac{\theta \varepsilon^2 \xi}{6 \lambda} \right)^{1/4}=\pm i^{3/2} i^{1/2} f^{1/4}=\mp f^{1/4}=\mp \left(- \frac{(\theta \varepsilon^2+\xi)}{6 \lambda} \right)^{1/4}. 
\end{equation}
These are two real solutions which are homogeneous in the sense that they depend only on the material parameters $\xi, \lambda, \theta$ and the loading constant $\varepsilon$. Now, eq. (61) renders $p_2=0$ when $\theta \neq \xi$. For this case eqs. (58), (59) are identically satisfied. So, collectively, the pair $(p_1, p_2)=(\left(- \frac{\theta \epsilon^2+\xi}{6 \lambda} \right)^{1/4}, 0)$ as well as the pair $(p_1, p_2)=(-\left(- \frac{\theta \epsilon^2+\xi}{6 \lambda} \right)^{1/4}, 0)$, qualifies as a solution to our problem for the case of axial tension/compression of the from of eq. (56). These are homogeneous solutions for the shift vector along the loading direction.

For the case when $\theta = - \xi$, eq. (61) is satisfied identically. Eq. (60) renders the five solutions of eq. (62), which only two are physically justifiable as above. Eqs. (58), (59) become, respectively
\begin{equation}
\varepsilon \theta p_1 p_{2,2}=0, \ \ \theta p_1 p_{2,1}+2 \theta p_2 p_{2,2}=0 
\end{equation}
which render $p_2=ct$, when $\theta, \varepsilon, p_1$ are non-zero. So, another two pairs that qualify as solutions are the pairs $(p_1, p_2)=(\left(- \frac{\theta \varepsilon^2+\xi}{6 \lambda} \right)^{1/4}, ct)$,  $(p_1, p_2)=(-\left(- \frac{\theta \varepsilon^2+\xi}{6 \lambda} \right)^{1/4}, ct)$ for the case when $\theta \neq - \xi$. These are homogeneous solutions for $\bf p$ along the loading direction which depend on the material parameters and the loading constant. On the other direction the shift vector equals to a constant to be determined by the boundary conditions. 

When $p_1=0$ and $\theta = -\xi$ eqs. (58), (60), (61) are satisfied trivially. So, eq. (59) renders
\begin{equation}
2 \theta p_2 p_{2,2}=0
\end{equation}
which is satisfied when $p_2=\phi(\Theta^1)$, $\phi$ being an arbitrary function. So, the pair $(p_1, p_2)=(0, \phi(\Theta^1))$ qualifies as a solution as well. This solution is not homogeneous and is along the direction perpendicular to the direction of the loading. The function $\phi$ should be determined by the boundary conditions. 

The components of the surface stress tensor for the pair $(p_1, p_2)=(\pm \left(- \frac{\theta \epsilon^2+\xi}{6 \lambda} \right)^{1/4}, ct)$ read
\begin{eqnarray}
&&S_{S _{11}}=\alpha + 3 \gamma \epsilon^4 \pm \theta \left(- \frac{\theta \epsilon^2+\xi}{6 \lambda} \right)^{1/2},  \nonumber\\
&&S_{S_{12}}=\pm \theta \left(- \frac{\theta \epsilon^2+\xi}{6 \lambda} \right)^{1/4} ct=S_{S_{21}}, \\
&&S_{S_{22}}=\alpha + \theta ct^2. \nonumber 
\end{eqnarray}
For the pair of solution $(p_1, p_2)=(\pm \left(- \frac{\theta \epsilon^2+\xi}{6 \lambda} \right)^{1/4}, 0)$ the stress components render
\begin{eqnarray}
&&S_{S _{11}}=\alpha + 3 \gamma \epsilon^4 \pm \theta \left(- \frac{\theta \epsilon^2+\xi}{6 \lambda} \right)^{1/2},  \nonumber\\
&&S_{S_{12}}=0=S_{S_{21}}, \\
&&S_{S_{22}}=\alpha. \nonumber 
\end{eqnarray}
Finally, for the case $(p_1, p_2)=(0, \phi(\Theta^1))$ the stress components are
\begin{eqnarray}
&&S_{S _{11}}=\alpha + 3 \gamma \epsilon^4,  \nonumber\\
&&S_{S_{12}}=0=S_{S_{21}}, \\
&&S_{S_{22}}=\alpha+\theta \phi^2(\Theta^1). \nonumber 
\end{eqnarray}
The traction vector that accompanies these stress tensors can be evaluated throught the relation $t_A=S_{S_{AB}} N_B$, $N_B$ being the outward unit normal to the current configuration.

Tension in the other direction is described by the deformation field
\begin{eqnarray}
&& x_1=\Theta^1 \nonumber\\
&& x_2=\varepsilon \Theta^2.
\end{eqnarray} 
The derivatives of the position vector in this case read
\begin{equation}
x_{1,1}=1, \ \ x_{1,2}=0, \ \ x_{2,2}=\varepsilon, \ \ x_{2,1}=0.
\end{equation} 
So, the equations (52-55) with the assumption of eq. (70) render
\begin{eqnarray}
&& 2 \theta p_1 p_{1,1} +  \theta p_{1,2} p_2+  \theta p_1 p_{2,2}=0, \\
&& \varepsilon \theta p_{1,1} p_2 +\theta p_1 p_{2,1} +2 \theta p_2 p_{2,2}=0, \\
&& \theta p_1 + 6\lambda p_1^5+\xi p_1=0 \rightarrow 6 \lambda p_1^5+(\theta +\xi) p_1=0, \\
&& \theta \varepsilon^2 p_2 + \xi p_2=0 \rightarrow (\theta \varepsilon^2+\xi) p_2=0.
\end{eqnarray}
By a reasoning similar as above, for the case when $\theta \varepsilon^2+\xi \neq 0$, $\frac{\kappa +\xi}{6 \lambda} < 0$ we have the following pair of solutions  $(p_1, p_2)=(\left(- \frac{\theta +\xi}{6 \lambda} \right)^{1/4}, 0)$,  $(p_1, p_2)=(-\left(- \frac{\theta +\xi}{6 \lambda} \right)^{1/4}, 0)$. The case $\theta \varepsilon^2+\xi = 0$, $\frac{\theta +\xi}{6 \lambda} < 0$ renders the following two pair of solutions  $(p_1, p_2)=(\left(- \frac{\theta +\xi}{6 \lambda} \right)^{1/4}, ct)$,  $(p_1, p_2)=(-\left(- \frac{\theta +\xi}{6 \lambda} \right)^{1/4}, ct)$. These are homogeneous solutions perpendicular to the loading direction while the constant should be determined by the boundary conditions. 

When $\theta \varepsilon^2+\xi=0$ and $p_1=0$ eqs. (72), (74), (75) are satisfied trivially. So, eq. (73) render 
\begin{equation}
2 \theta p_2 p_{2,1}=0
\end{equation}
which is satisfied when $p_2=\phi(\Theta^1)$, $\phi$ being an arbitrary function. So, the pair $(p_1, p_2)=(0, \phi(\Theta^1))$ qualifies as a solution to this case. This is an inhomogeneous solution along the loading direction and $\phi$ should be determined by the boundary conditions. The components of the stress tensor for the pair  $(p_1, p_2)=( \pm \left(- \frac{\theta +\xi}{6 \lambda} \right)^{1/4}, ct)$ read
\begin{eqnarray}
&&S_{S _{11}}=\alpha + 3 \gamma  \pm \theta  \left(- \frac{\theta +\xi}{6 \lambda} \right)^{1/2},  \nonumber\\
&&S_{S_{12}}=\pm \theta \left(- \frac{\theta +\xi}{6 \lambda} \right)^{1/4} ct=S_{S_{21}}, \\
&&S_{S_{22}}=\alpha + \theta ct^2. \nonumber 
\end{eqnarray}
For the pair $(p_1, p_2)=( \pm \left(- \frac{\theta +\xi}{6 \lambda} \right)^{1/4}, 0)$ we obtain 
\begin{eqnarray}
&&S_{S _{11}}=\alpha + 3 \gamma  \pm \theta  \left(- \frac{\theta +\xi}{6 \lambda} \right)^{1/2},  \nonumber\\
&&S_{S_{12}}=0=S_{S_{21}}, \\
&&S_{S_{22}}=\alpha. \nonumber 
\end{eqnarray}
Finally, the pair $(p_1, p_2)=(0, \phi(\Theta^1))$ gives for the stress tensor
\begin{eqnarray}
&&S_{S _{11}}=\alpha + 3 \gamma,  \nonumber\\
&&S_{S_{12}}=0=S_{S_{21}}, \\
&&S_{S_{22}}=\alpha + \theta \phi^2(\Theta^1). \nonumber 
\end{eqnarray}

The above analysis shows that in the case of one dimensional tensile/compressive mechanical loading of the graphene sheet, analytical solutions of the problem are feasible for the components of the shift vector. These solutions are the following pairs: a) the pair of trivial solutions, b) the pair consisting of a homogeneous solution depending on the material parameters along the direction of loading while on the other direction it can be zero, or constant and c) a pair consisting of a trivial solution at the direction of loading with a generic function depending on one independent variable along the other direction. When the loading changes direction, the solutions obtained are similar in form with the ones of the previous case. However, now the homogeneous solution is perpendicular to the loading direction while the non-homogeneous solution is along the loading direction.  

\subsection{Biaxial tension/compression}

Biaxial loading is described by  
\begin{equation}
x_1 = \varepsilon_1 \Theta^1, \ \ x_2=\varepsilon_2 \Theta^2,
\end{equation}
thus giving for the derivatives
\begin{equation}
x_{1,1}=\varepsilon_1, \ \ x_{2,2}=\varepsilon_2, \ \ x_{1,2}=x_{2,1}=0. 
\end{equation}

The field equations now take the form 
\begin{eqnarray}
&& \varepsilon_1 2 \theta p_{1,1} p_1+\varepsilon_1 \theta p_{1,2} p_2+\varepsilon_1 \theta p_1 p_{2,2} =0, \\
&& \varepsilon_2 \theta p_{1,1} p_2 +\varepsilon_2 \theta p_1 p_{2,1}+\varepsilon_2 \theta 2 p_{2,2} p_2 =0, \\
&& \theta \varepsilon_1^2 p_1 + 6 \lambda p_1^5 +\xi p_1 =0, \\
&& \theta \varepsilon_2^2 p_2 +\xi p_2=0.
\end{eqnarray}
The case when $\theta \varepsilon_2^2+\xi \neq 0$, $\frac{\theta \varepsilon_1^2+\xi}{6 \lambda} < 0 $ render the pair of solutions  $(p_1, p_2)=(\left(- \frac{\theta \varepsilon_1^2 +\xi}{6 \lambda} \right)^{1/4}, 0)$,  $(p_1, p_2)=(-\left(- \frac{\theta \varepsilon_1^2+\xi}{6 \lambda} \right)^{1/4}, 0)$. The case $\theta \varepsilon_2^2 + \xi=0 $ renders from eq. (85) that $p_1=0$, so eq. (82) is satisfied trivially, while eq. (83) gives $\varepsilon_2 \theta p_2 p_{2,2}=0 \rightarrow p_2=\phi (\Theta^1)$. Collectively, for this case the pair $(p_1, p_2)=(0, \phi (\Theta^1))$ qualifies as a solution to our problem. 

The components of the stress tensor for the pair $(p_1, p_2)=( \pm \left(- \frac{\theta \varepsilon_1^2 +\xi}{6 \lambda} \right)^{1/4}, 0)$ read
\begin{eqnarray}
&&S_{S _{11}}=\alpha + 3 \gamma \varepsilon_1^4 \pm \theta  \left(- \frac{\theta \varepsilon_1^2 +\xi}{6 \lambda} \right)^{1/2},  \nonumber\\
&&S_{S_{12}}=0=S_{S_{21}}, \\
&&S_{S_{22}}=\alpha. \nonumber 
\end{eqnarray}
For the pair $(p_1, p_2)=(0, \phi (\Theta^1))$ we obtain 
\begin{eqnarray}
&&S_{S _{11}}=\alpha + 3 \gamma  \varepsilon_1^4,  \nonumber\\
&&S_{S_{12}}=0=S_{S_{21}}, \\
&&S_{S_{22}}=\alpha+\theta \phi^2(\Theta^1). \nonumber 
\end{eqnarray}

The biaxial mechanical loading, which is a more general case of loading for graphene, can also be treated analytically, like the simple tension/compression case. As in the previous case the solutions are pairs: a) the pair of trivial solutions, b) the pair consisting of a homogeneous solution depending on the material parameters while on the other direction it can be zero, or constant and c) the pair consisting of a trivial solution with a generic function depending on one independent variable along the other direction. Compared to the corresponding solutions of the one dimensional tension/compression case, the solutions have the same form, but different material parameters are present in their expressions. 

\subsection{Simple shear}

Another interesting mechanical loading we examine is the simple shear depicted in Figure 6
\begin{figure}[!htb]
\centering
\includegraphics{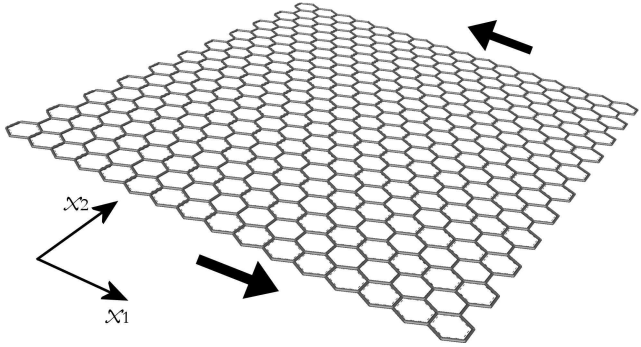}
\caption{In-plane simple shear.}
\label{fig:digraph}
\end{figure}
and described by  
\begin{equation}
x_1=\Theta^1+\varepsilon \Theta^2, \ \ x_2=\Theta^2,
\end{equation}
so for the necessary derivatives we have
\begin{equation}
x_{1,1}=1, \ \ x_{1,2}=\varepsilon, \ \ x_{2,1}=0, \ \ x_{2,2}=1.
\end{equation}
The field equations take the form 
\begin{eqnarray}
&& [\alpha +\theta p_1^2]_{,1}+[\theta p_1 p_2]_{,2}+[\epsilon \theta p_1 p_2]_{,1}+[\epsilon (\alpha +\theta p_2^2)]_{,2}=0, \\
&& [\theta p_1 p_2]_{,1}+[\alpha + \theta p_2^2]_{,2}=0, \\
&& \theta p_1 +\theta \varepsilon p_2 +6 \lambda p_1^5+\xi p_1=0, \\
&& \theta \varepsilon p_1+ \theta (\varepsilon^2+1) p_2 + \xi p_2=0.
\end{eqnarray}

When $\theta \varepsilon^2+\theta +\xi \neq 0$, from eq. (93) we obtain $p_2=\frac{- \theta \varepsilon}{\theta \varepsilon^2+\kappa+\xi} p_1$ which when substituted in eq. (92) renders
\begin{equation}
6 \lambda p_1^5+\left( \theta+\xi -\frac{\theta^2 \varepsilon^2}{\theta \varepsilon^2+\theta +\xi}  \right) p_1=0.
\end{equation}
For the case when $\left( \theta + \xi -\frac{\theta^2 \varepsilon^2}{\theta \varepsilon^2+\theta +\xi}  \right) \lambda < 0$ we obtain the following pair of solutions $(p_1, p_2)=(\left(- \frac{ \theta + \xi -\frac{\theta^2 \varepsilon^2}{\theta \varepsilon^2+\theta +\xi}}{6 \lambda} \right)^{1/4},\frac{\theta \varepsilon}{\theta \varepsilon^2+\theta + \xi}\left(- \frac{ \theta + \xi -\frac{\theta^2 \varepsilon^2}{\theta \varepsilon^2 + \theta + \xi}}{6 \lambda} \right)^{1/4} )$, \\ $(p_1, p_2)=(-\left(- \frac{ \theta + \xi -\frac{\theta^2 \varepsilon^2}{\theta \varepsilon^2+\theta +\xi}}{6 \lambda} \right)^{1/4}, -\frac{\theta \varepsilon}{\theta \varepsilon^2 + \theta + xi} \left(- \frac{ \theta + \xi -\frac{\theta^2 \varepsilon^2}{\theta \varepsilon^2+\theta +\xi}}{6 \lambda} \right)^{1/4} )$. These solutions are homogeneous in the sense that they depend on the material parameters $\theta, \xi, \lambda$ and the loading constant $\varepsilon$.

When $\theta \varepsilon^2 +\theta +\xi =0$ eq. (93) gives $p_1=0$ and eq. (92) $p_2=0$. So, for this case the trivial pair qualifies as a solution for the problem at hand. Collectively, we have the trivial solution for the shift vector and a homogeneous solution along both directions depending on the material parameters. Stress components can be calculated in a similar fashion as in the previous sections.  

${\bf{Remark}}$\\
Comparing the tension/compression cases with those of the simple shear there seems to be some lack of symmetry in the results of $p_1, p_2$. This is due to the term $x_{1,2}$ which is zero for the tension/compression case, while it is non-zero for the simple shear problem. As an outcome of that, eqs. (61, 75) contain only $p_2$ while eq. (93) contains both $p_1, p_2$.

\section{Out-of-plane deformations of a monolayer graphene}

When out-of-plane motions are taken into account, the $x_3$ component of the position vector $\bf x$ is no longer constant; it describes the out of plane motion of the surface. In this case, the first of the momentum equation acquires the form
\begin{eqnarray}
&& \{ x_{1,1} [ \alpha +3 \gamma (x_{1,1}^2+x_{2,1}^2+x_{3,1}^2)^2 + \theta p_1^2  ] \}_{,1} +\{x_{1,1} [\theta p_1 p_2 ]   \}_{,2} + \{x_{1,2} [\theta p_1 p_2 ]   \}_{,1} \nonumber\\
&&+\{ x_{1,2} [\alpha +\theta p_2^2]   \}_{,2}=0. 
\end{eqnarray}
The second of the momentum equation renders
\begin{eqnarray}
&& \{ x_{2,1}  [ \alpha +3 \gamma (x_{1,1}^2+x_{2,1}^2+x_{3,1}^2)^2 +\theta p_1^2   ]    \}_{,1}  +\{ x_{2,1}  [ \theta p_1 p_2    ]     \}_{,2} + \{ x_{2,2} [\theta p_1 p_2]    \}_{,1} \nonumber\\
&& +\{ x_{2,2} [ \alpha +\theta p_2^2   ]   \}_{,2}=0. 
\end{eqnarray}
For the auxiliary variables the first equation reads
\begin{eqnarray}
&& \theta (x_{1,1}^2+x_{2,1}^2+x_{3,1}^2) p_1 + \theta (x_{1,1}x_{1,2}+x_{2,1}x_{2,2}+x_{3,1}x_{3,2}) p_2 \nonumber\\
&& +\iota b_{11} p_1 + \iota b_{12} p_2 +6 \lambda p_1^5+ \xi p_1=0.
\end{eqnarray}
The second equation of the auxiliary variables reads
\begin{eqnarray}
&& \theta (x_{1,1} x_{1,2} + x_{2,1}x_{2,2}+x_{3,1}x_{3,2}) p_1 +\theta (x_{1,2}^2+x_{2,2}^2+x_{3,2}^2) p_2 \nonumber\\
&& +\iota b_{21} p_1 +\iota b_{22} p_2 +\xi p_2=0.
\end{eqnarray}

For the first equation of the moment of momentum we have
\begin{eqnarray}
&& \{ x_{1,1} [\epsilon+3 \eta (x_{1,1}^2+x_{2,1}^2+x_{3,1}^2)^2]+\iota p_1^2      \}_{,1}  +\{ x_{1,2} [\iota p_1 p_2]   \}_{,1} + \{ x_{1,1} [\iota p_1 p_2]   \}_{,2} \nonumber\\
&& +\{ x_{1,2} [\epsilon + \iota p_2^2] \}_{,2}  -x_{1,1} \{ x_{1,1} [ \theta p_1 p_2]+x_{1,2} [\alpha +\theta p_2^2]    \} \nonumber\\
&& -x_{1,2} \{ x_{1,1} [ \theta p_1 p_2]+x_{1,2} [\alpha +\theta p_2^2]    \} \nonumber\\
&& +x_{2,1} \{ x_{1,1} [\alpha +3 \gamma (x_{1,1}^2+x_{2,1}^2+x_{3,1}^2)^2 + \theta p_1^2]+x_{1,2} [\theta p_1 p_2]    \} \nonumber\\
&& +x_{2,2} \{ x_{1,1} [\alpha +3 \gamma (x_{1,1}^2+x_{2,1}^2+x_{3,1}^2)^2 + \theta p_1^2]+x_{1,2} [\theta p_1 p_2]    \} \nonumber\\
&& -x_1 \{ x_{1,1}[\theta p_1 p_2]+x_{1,2}[\alpha +\theta p_2^2]    \}_{,1}  -x_1 \{ x_{1,1}[\theta p_1 p_2]+x_{1,2}[\alpha +\theta p_2^2]    \}_{,2} \nonumber\\
&& +x_2 \{ x_{1,1} [\alpha +3 \gamma (x_{1,1}^2+x_{2,1}^2+x_{3,1}^2)^2 +\theta p_1^2  ]  + x_{1,2} [\theta p_1 p_2]     \}_{,1} \nonumber\\
&& +x_2 \{ x_{1,1} [\alpha +3 \gamma (x_{1,1}^2+x_{2,1}^2+x_{3,1}^2)^2 +\theta p_1^2  ]  + x_{1,2} [\theta p_1 p_2]     \}_{,2}=0.
\end{eqnarray}
The second of the moment of momentum equations reads
\begin{eqnarray}
&& \{ x_{2,1} [\epsilon +3 \eta (x_{1,1}^2+x_{2,1}^2+x_{3,1}^2)^2+\iota p_1^2 ]   \}_{,1}  +\{ x_{2,2} [\iota p_1 p_2]   \}_{,1} + \{ x_{2,1} [\iota p_1 p_2 ] \}_{,2} \nonumber\\
&& +\{ x_{2,2} [\epsilon +\iota p_2^2] \}_{,2}  -x_{1,1} \{ x_{2,1} [\theta p_1p_2]+x_{2,2}[\alpha + \theta p_2^2]     \} \nonumber\\
&& -x_{1,2}  \{ x_{2,1} [\theta p_1p_2]+x_{2,2}[\alpha + \theta p_2^2]     \} \nonumber\\
&& +x_{2,1} \{ x_{2,1} [\alpha +3 \gamma (x_{1,1}^2+x_{2,1}^2+x_{3,1}^2)^2+\theta p_1^2 ]+x_{2,2} [\theta p_1 p_2]    \} \nonumber\\
&& +x_{2,2} \{ x_{2,1} [\alpha +3 \gamma (x_{1,1}^2+x_{2,1}^2+x_{3,1}^2)^2+\theta p_1^2 ]+x_{2,2} [\theta p_1 p_2]    \} \nonumber\\
&& -x_1 \{x_{2,1} [\theta p_1 p_2] +x_{2,2} [\alpha +\theta p_2^2]     \}_{,1}  -x_1 \{x_{2,1} [\theta p_1 p_2] +x_{2,2} [\alpha +\theta p_2^2]     \}_{,2} \nonumber\\
&& +x_2 \{x_{2,1} [\alpha +3 \gamma (x_{1,1}^2+x_{2,1}^2+x_{3,1}^2)^2+\theta p_1^2]  +x_{2,2}[\theta p_1 p_2]    \}_{,1} \nonumber\\
&& +x_2 \{x_{2,1} [\alpha +3 \gamma (x_{1,1}^2+x_{2,1}^2+x_{3,1}^2)^2+\theta p_1^2]  +x_{2,2}[\theta p_1 p_2]    \}_{,2}=0
\end{eqnarray}
The above equations differ from the corresponding ones for in plane motions, at three levels: a. at the momentum equation the coordinate $x_3$ is also present, b. at the equations ruling the auxiliary variables the components of the curvature are present (thereby making them unable to be solved, in contrast to the motion in plane), c. the equations of moment of momentum are also present. The mathematical analysis of this problem is much more complicated compared to the previous cases, but enables us to study more complex phenomena such as wrinkling and expand the investigation for the mechanical loadings we studied in previous sections. 

\subsection{Introducing wrinkling/buckling}
In order to model wrinkling/buckling we need to assume that the out of plane displacement is given by the following expression (\cite{Timoshenko,Punteletal2011})
\begin{equation}
x_3=x_3(\Theta^1, \Theta^2)=cos \left( \frac{n \pi \Theta^1}{2 L_1}   \right) f(\Theta^2), 
\end{equation}
$n$ being the number of sinusoidal wave in the $\Theta^1$ direction and $f$ is an arbitrary function (see Figure 7
\begin{figure}[!htb]
\centering
\includegraphics{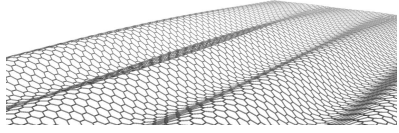}
\caption{Wrinkling/buckling described by eq. (101) (figure taken from \cite{Androulidakisetal2014}).}
\label{fig:digraph}
\end{figure}
for a schematic guide for this kind of deformation). The parametric form of a surface having the above expression as displacement is 
\begin{equation}
{\bf x}(\Theta^1, \Theta^2)=(\Theta^1, \Theta^2, cos \left( \frac{n \pi \Theta^1}{2 L_1}  \right) f(\Theta^2)).
\end{equation}
For our framework, $\bf b$ is the second fundamental form of the surface, so we evaluate for its components
\begin{eqnarray}
&& b_{11}={\bf x}_{,11} \cdot {\bf n} \\
&& b_{12}=b_{21}={\bf x}_{,12} \cdot {\bf n} \\
&& b_{22}={\bf x}_{,22} \cdot {\bf n}.
\end{eqnarray}
The outward unit normal of the surface is defined as usual by 
\begin{equation}
{\bf n}=\frac{{\bf x}_{,1} \times  {\bf x}_{,2}  }{|{\bf x}_{,1} \times  {\bf x}_{,2}| }.
\end{equation}
These measures of the surface are important since they participate to the field equations (95-100) when out-of-plane motions are taken into acoount.

\subsection{Tension/Compression}

Axial tension/compression resulting in wrinkling/buckling is described by the parametric form of the surface 
\begin{equation}
{\bf x}(\Theta^1, \Theta^2)= \left( \varepsilon \Theta^1, \Theta^2, \textrm{cos} \left( \frac{n \pi \Theta^1}{2 L_1} \right) f(\Theta^2) \right).
\end{equation}
Such an assumption means that tension/compression in the in-plane results in wrinkling/buckling, i.e. out of plane motion. The phenomenon is not assumed to be dynamic in order to have tension/compression initially that finally leads to wrinkling/buckling. The method is semi-inverse: we assume the form that the solution has in the final form. Tension will finally produce wrinkling on the material, while compression will lead to buckling. Certainly, one expects different behaviour in these two kind of loadings. Such a hardening response cannot be captured by the model in its present form; generalizations should be made which are outside the scope of this work. 

For the above given surface the outward unit normal has components
\begin{equation}
{\bf n}= \left( -\frac{n \pi}{2 L_1} \textrm{sin}   \left( \frac{n \pi \Theta^1}{2 L_1} \right) f(\Theta^2), -\varepsilon \textrm{cos}   \left( \frac{n \pi \Theta^1}{2 L_1} \right) f'(\Theta^2), \varepsilon \right),
\end{equation}
when for its Euclidean length we assume it is unity: 
\begin{equation}
||{\bf n}||=\sqrt{\left[ -\frac{n \pi}{2 L_1} \textrm{sin}   \left( \frac{n \pi \Theta^1}{2 L_1} \right) f(\Theta^2) \right]^2+\left[ -\varepsilon \textrm{cos}   \left( \frac{n \pi \Theta^1}{2 L_1} \right) f'(\Theta^2) \right]^2+\varepsilon^2}=1.
\end{equation}
For the components of the second fundamental form we then obtain
\begin{eqnarray}
&& b_{11}=-\varepsilon \frac{n^2 \pi^2}{4 L_1^2} \textrm{cos}   \left( \frac{n \pi \Theta^1}{2 L_1} \right) f(\Theta^2), \\
&& b_{12}=b_{21}=-\varepsilon \frac{n \pi}{2 L_1} \textrm{sin}   \left( \frac{n \pi \Theta^1}{2 L_1} \right) f'(\Theta^2), \\
&& b_{22}=\varepsilon \textrm{cos}   \left( \frac{n \pi \Theta^1}{2 L_1} \right) f''(\Theta^2).
\end{eqnarray}

The first of the momentum equations then reads
\begin{eqnarray}
&& \{ \varepsilon [ \alpha +3 \gamma \left( \varepsilon^2+ (-\varepsilon \frac{n \pi}{2 L_1} \textrm{sin}   \left( \frac{n \pi \Theta^1}{2 L_1}) \right) f'(\Theta^2))^2 \right)^2 +\theta p_1^2 ]    \}_{,1}  +\{ \varepsilon \theta p_1 p_2  \}_{,2}=0.
\end{eqnarray}
The second of the momentum equations reads
\begin{equation}
\{ \theta p_1 p_2  \}_{,1} + \{ \alpha +\theta p_2^2  \}_{,2}=0.
\end{equation}
For the auxiliary variables the first equation reads
\begin{eqnarray}
&& \theta \left[ \varepsilon^2 + \left[- \frac{n \pi}{2 L_1} \textrm{sin}   \left( \frac{n \pi \Theta^1}{2 L_1} \right) f(\Theta^2)  \right]^2 \right] p_1 \nonumber\\
&& -\theta \frac{n \pi}{2 L_1} \textrm{sin}   \left( \frac{n \pi \Theta^1}{2 L_1} \right) f(\Theta^2) \frac{n \pi}{2 L_1} \textrm{cos}   \left( \frac{n \pi \Theta^1}{2 L_1} \right) f'(\Theta^2) p_2 \nonumber\\
&& -\iota \varepsilon \frac{n^2 \pi^2}{4 L_1^2} \textrm{cos}   \left( \frac{n \pi \Theta^1}{2 L_1} \right) f(\Theta^2) p_1  -\iota \varepsilon \frac{n \pi}{2 L_1} \textrm{sin}   \left( \frac{n \pi \Theta^1}{2 L_1} \right) f'(\Theta^2) p_1 \nonumber\\
&& +6 \lambda p_1^5+\xi p_1=0, 
\end{eqnarray}
while the second equation ruling the auxiliary variables reads
\begin{eqnarray}
&& -\theta  \frac{n \pi}{2 L_1} \textrm{sin}   \left( \frac{n \pi \Theta^1}{2 L_1} \right) f(\Theta^2) \textrm{cos}   \left( \frac{n \pi \Theta^1}{2 L_1} \right) f(\Theta^2) p_1 \nonumber\\
&& \theta \left[1+ \left( \textrm{cos}   \left( \frac{n \pi \Theta^1}{2 L_1} \right) f'(\Theta^2) \right)^2 \right]p_2 -\iota \varepsilon  \frac{n \pi}{2 L_1} \textrm{sin}   \left( \frac{n \pi \Theta^1}{2 L_1} \right) f'(\Theta^2) p_1 \nonumber\\
&& +\iota \varepsilon  \textrm{sin}   \left( \frac{n \pi \Theta^1}{2 L_1} \right) f''(\Theta^2) p_2  +\xi p_2=0.
\end{eqnarray}

The first equation of moment of momentum renders
\begin{eqnarray}
&& \{ \varepsilon [\epsilon +3 \eta \left[ \varepsilon^2+\left(-\frac{n \pi }{2 L_1} \textrm{sin} \left( \frac{n \pi \Theta^1}{2 L_1} \right) f(\Theta^2)  \right)^2 \right]^2  +\iota p_1^2   ]   \}_{,1} \nonumber\\
&& +\{ \varepsilon [\iota p_1 p_2]  \}_{,2}-\varepsilon \{ \varepsilon [ \theta p_1 p_2 ]   \} \nonumber\\
&& +\{ \varepsilon [\alpha +3 \gamma \left( \varepsilon^2+  \left[ \frac{n \pi }{2 L_1} \textrm{sin} \left( \frac{n \pi \Theta^1}{2 L_1} \right) f(\Theta^2) \right]^2  \right)^2   +\theta p_1^2  ]   \} \nonumber\\
&& -\varepsilon \Theta^1 \{ \varepsilon [\theta p_1 p_2]   \}_{,1} -\varepsilon \Theta^1 \{ \varepsilon [\theta p_1 p_2]    \}_{,2} \nonumber\\
&& +\Theta^2 \{ \varepsilon [\alpha + 3 \gamma \left( \varepsilon^2+  \left[ \frac{n \pi }{2 L_1} \textrm{sin} \left( \frac{n \pi \Theta^1}{2 L_1} \right) f(\Theta^2) \right]^2  \right)^2 +\theta p_1^2  ]     \}_{,1} \nonumber\\
&&  +\Theta^2 \{ \varepsilon [\alpha + 3 \gamma \left( \varepsilon^2+  \left[ \frac{n \pi }{2 L_1} \textrm{sin} \left( \frac{n \pi \Theta^1}{2 L_1} \right) f(\Theta^2) \right]^2  \right)^2 +\theta p_1^2  ]     \}_{,2}=0 
\end{eqnarray}
and the second becomes
\begin{eqnarray}
&& \{ [ \iota p_1 p_2]     \}_{,1} + \{ \epsilon +\iota p_2^2     \}_{,2} - \varepsilon \{ [\alpha +\theta p_2^2]   \} + \{ \theta p_1 p_2  \} \nonumber\\
&& -\varepsilon \Theta^1 \{ \alpha +  \theta p_2^2    \}_{,1}  - \varepsilon \Theta^2 \{ \alpha +  \theta p_2^2  \}_{,2}+\Theta^2 \{ \theta p_1 p_2  \}_{,1} +\Theta^2 \{ \theta p_1 p_2   \}_{,2}=0
\end{eqnarray}

For the momentum equations we therefore have
\begin{eqnarray}
&& (2 \varepsilon \theta p_1) p_{1,1}+(\varepsilon \theta p_2) p_{1,2}+\ \ \ \ \ \ \ \ \ \ \ \ + (\varepsilon \theta p_1)p_{2,2}=\mathcal G, \\
&& (\theta p_2) p_{1,1} \ \ \ \ \ \ \ \ \ \ \ \ \ \ \ \ \ \ \ +(\theta p_1) p_{2,1}  +(2 \theta p_2)p_{2,2}=0, 
\end{eqnarray}
where for the coefficient $\mathcal G$ it holds
\begin{eqnarray}
\mathcal G=&&12 \varepsilon^3 \gamma \left( \frac{n \pi }{2 L_1} \right)^3 \textrm{sin} \left( \frac{n \pi \Theta^1}{2 L_1} \right)  \textrm{cos} \left( \frac{n \pi \Theta^1}{2 L_1} \right) f(\Theta^2)^2 \nonumber\\
&&+3  \left( \frac{n \pi }{2 L_1} \right)^5 \textrm{sin}^3 \left( \frac{n \pi \Theta^1}{2 L_1} \right)  \textrm{cos} \left( \frac{n \pi \Theta^1}{2 L_1} \right) f(\Theta^2)^4.
\end{eqnarray}
So, when eqs. (119-120) are viewed as a quasilinear system in terms of $(p_1, p_2)$ we take
\begin{equation}
{\bf A}= \left(\begin{array}{cc}
2 \varepsilon \theta p_1 &  \varepsilon \theta p_2     \\
\theta p_2 & 0
\end{array}\right), 
{\bf B}= \left(\begin{array}{cc}
0 &  \varepsilon \theta p_1     \\
\theta p_1 & 2 \theta p_2
\end{array}\right),
{\bf C}= \left(\begin{array}{c}
\mathcal G    \\
0
\end{array}\right).
\end{equation}

Cauchy data specify $\bf p$ on a curve $\Gamma$ in the $(\Theta^1, \Theta^2)$ plane, i.e. 
\begin{equation}
\Theta^1=x_0(s), \ \ \Theta^2=y_0(s), \ \ {\bf p}={\bf p}_0(s).
\end{equation}
The condition of the initial data for the first derivative of $\bf p$ to be locally determined reads
\begin{equation}
\textrm{det} \left(\begin{array}{cccc}
2 \varepsilon \theta p_1 &  \varepsilon \theta p_2 & 0 & \varepsilon \theta p_1     \\
\theta p_2 & 0 & \theta p_1 & 2 \theta p_2 \\
x'_0 & 0 & y'_0  & 0 \\
0 & x'_0 & 0 & y'_0
\end{array}\right) \neq 0.
\end{equation}
This condition further simplifies to
\begin{eqnarray}
&& -\varepsilon \theta^2 p_1^2 (x'_0)^2-2 \varepsilon \theta^2 p_1 p_2 x'_0 y'_0 -\varepsilon \theta^2 p_2^2  (y'_0)^2 \neq 0 
\end{eqnarray}
When 
\begin{equation}
\textrm{det} {\bf A} \neq 0 \Rightarrow \varepsilon \theta^2 p_1^2 \neq 0
\end{equation}
the Cauchy-Kowalevski theorem renders existence and uniqueness for an analytic function ${\bf p}={\bf p}(\Theta^1, \Theta^2)$ in a neighborhood of $(0, y_0)$ for the system of eqs. (119), (120) when $\mathcal G$ ia analytic in its arguments. The eigenvalues for this problem read
\begin{equation}
\lambda_{1,2}=-\frac{p_1}{p_2},
\end{equation}
so we speak about a parabolic problem, since we have a double real eigenvalue.

\section{General approach }

The full problem is very difficult to treat. There is no assumption that reduces the number of material parameters so the field equations become really lengthy. We derive the momentum equations and classify them for the case of axial tension/compression with wrinkling/buckling, namely eq. (107). For the momentum equations we have
\begin{eqnarray}
&& 2 \varepsilon \theta p_1 p_{1,1}+4 \varepsilon \varrho p_1^3 p_{1,1}+\varepsilon \theta p_{1,2} p_2+\varepsilon \theta p_1 p_{2,2}=-\mathcal P, \\
&& \theta p_2 p_{1,1}+\theta p_1 p_{2,1}+2 \theta p_2 p_{2,2}=-\mathcal Q, 
\end{eqnarray}
where $\mathcal P$, $\mathcal Q$ are given by 
\begin{eqnarray}
\mathcal P=&&-2 \varepsilon \beta \frac{n \pi }{2 L_1} \textrm{sin} \left(  \frac{n \pi \Theta^1}{2 L_1}    \right) f(\Theta^2)-6 \gamma \varepsilon^2   \left( \frac{n \pi }{2 L_1} \right)^2 \textrm{cos} \left(  \frac{n \pi \Theta^1}{2 L_1}    \right) f(\Theta^2) \nonumber\\
&&-3 \gamma   \left( \frac{n \pi }{2 L_1} \right)^2  \textrm{sin} \left(  \frac{n \pi \Theta^1}{2 L_1} \right)    \textrm{cos} \left(  \frac{n \pi \Theta^1}{2 L_1}    \right) f(\Theta^2)^2 +\varepsilon \delta  \left( \frac{n \pi }{2 L_1} \right)^3 \textrm{sin} \left(  \frac{n \pi \Theta^1}{2 L_1}    \right) f(\Theta^2) \nonumber\\
&&+\varepsilon \beta  \left( \frac{n \pi }{2 L_1} \right)^3  \textrm{sin} \left(  \frac{n \pi \Theta^1}{2 L_1} \right)    \textrm{cos} \left(  \frac{n \pi \Theta^1}{2 L_1}    \right) f'(\Theta^2)^2  \\
&&+\varepsilon \beta  \left( \frac{n \pi }{2 L_1} \right)  \textrm{sin} \left(  \frac{n \pi \Theta^1}{2 L_1} \right)    \textrm{cos} \left(  \frac{n \pi \Theta^1}{2 L_1}    \right) f(\Theta^2) f''(\Theta^2)^2 -\delta \varepsilon  \left( \frac{n \pi }{2 L_1} \right) \textrm{sin} \left(  \frac{n \pi \Theta^1}{2 L_1}    \right) f''(\Theta^2), \nonumber
\end{eqnarray}
\begin{eqnarray}
\mathcal Q=&&\beta \left( \frac{n \pi }{2 L_1} \right)^2 \textrm{cos}^2 \left(  \frac{n \pi \Theta^1}{2 L_1}    \right) f(\Theta^2)  f'(\Theta^2)  -\beta  \left( \frac{n \pi }{2 L_1} \right)^2 \textrm{sin}^2 \left(  \frac{n \pi \Theta^1}{2 L_1}    \right) f(\Theta^2)  f'(\Theta^2) \nonumber\\
&&-\delta \varepsilon  \left( \frac{n \pi }{2 L_1} \right)^2 \textrm{cos} \left(  \frac{n \pi \Theta^1}{2 L_1}    \right) f'(\Theta^2) -2 \beta \varepsilon  \left( \frac{n \pi }{2 L_1} \right) \textrm{sin} \left(  \frac{n \pi \Theta^1}{2 L_1}    \right) f'(\Theta^2) \nonumber\\
&&+4 \beta \varepsilon  \left( \frac{n \pi }{2 L_1} \right) \textrm{sin}^2 \left(  \frac{n \pi \Theta^1}{2 L_1}    \right) f(\Theta^2) f'(\Theta^2) +\delta \varepsilon   \textrm{cos} \left(  \frac{n \pi \Theta^1}{2 L_1}    \right) f'''(\Theta^2).
\end{eqnarray}

When we view eqs. (128-129) as a quasilinear system we have for the relevant matrices  
\begin{equation}
{\bf A}= \left(\begin{array}{cc}
2 \varepsilon \theta p_1+4 \varepsilon \varrho p_1^3 &  \varepsilon \theta p_2     \\
\theta p_2 & 0
\end{array}\right), 
{\bf B}= \left(\begin{array}{cc}
0 &  \varepsilon \theta p_1     \\
\theta p_1 & 2 \theta p_2
\end{array}\right),
{\bf C}= \left(\begin{array}{c}
\mathcal P    \\
\mathcal Q
\end{array}\right).
\end{equation}

Cauchy data specify $\bf p$ on a curve $\Gamma$ in the $(\Theta^1, \Theta^2)$ plane, i.e. 
\begin{equation}
\Theta^1=x_0(s), \ \ \Theta^2=y_0(s), \ \ {\bf p}={\bf p}_0(s).
\end{equation}
The condition of the initial data for the first derivative of $\bf p$ to be locally determined reads
\begin{equation}
\textrm{det} \left(\begin{array}{cccc}
2 \varepsilon \theta p_1+4 \varepsilon \varrho p_1^3 &  \varepsilon \theta p_2 & 0 & \varepsilon \theta p_1     \\
\theta p_2 & 0 & \theta p_1 & 2 \theta p_2 \\
x'_0 & 0 & y'_0  & 0 \\
0 & x'_0 & 0 & y'_0
\end{array}\right) \neq 0.
\end{equation}
this condition further simplifies to
\begin{eqnarray}
&& -\varepsilon \theta^2 p_1^2 (x'_0)^2-2 \varepsilon \theta^2 p_1 p_2 x'_0 y'_0 -8 \varepsilon \theta \varrho p_1^3 p_2 x'_0 y'_0    -\varepsilon \theta^2 p_2^2  (y'_0)^2 \neq 0 
\end{eqnarray}
When 
\begin{equation}
\textrm{det} {\bf A} \neq 0 \Rightarrow \varepsilon \theta^2 p_2 \neq 0
\end{equation}
the Cauchy-kowalevski theorem renders existence and uniqueness for an analytic function ${\bf p}={\bf p}(\Theta^1, \Theta^2)$ in a neighborhood of $(0, y_0)$ for the system of eqs. (128), (129) when $\mathcal P, \mathcal Q$ are analytic in their arguments.

Eigenvalues for this case read
\begin{equation}
\lambda_{1,2}=\frac{-\theta p_1 p_2 -4 \varrho p_1^3 p_2 \pm 2 \sqrt{2}  \sqrt{\theta \varrho p_1^4 p_2^2+2 \varrho^2 p_1^6 p_2^2}  }{\theta p_2^2}.
\end{equation}
Depending on the behaviour of the term $\theta \varrho p_1^4 p_2^2+2 \varrho^2 p_1^6 p_2^2$ we get three different cases. Firslty, the case $\theta \varrho p_1^4 p_2^2+2 \varrho^2 p_1^6 p_2^2 > 0$ render two real eigenvalus thereby we speak about a hyperbolic problem. Secondly, the case $\theta \varrho p_1^4 p_2^2+2 \varrho^2 p_1^6 p_2^2 < 0$ redenr two complex eigenvalues so we speak about an elliptic problem. Finally, the case $\theta \varrho p_1^4 p_2^2+2 \varrho^2 p_1^6 p_2^2 = 0$ render a double real eigenvalues so we speak about a parabolic problem.

\section{Conclusion and future directions }

This work constitutes an extension of \cite{Sfyris-GaliotisMMS} in the direction of giving some closed form solutions for a free standing monolayer graphene. The approach is valid for geometrical as well as material nonlinearities at the level of the continuum. 

For the case of in plane motions we examine one dimensional tension/compression along both directions of the surface as well as the case of biaxial tension/compression and simple shear. The outcome consists of homogeneous solutions for the components of the shift vector that depend on the material parameters and the loading constant. For modeling out of plane motions we describe how wrinkling/buckling can be introduced into the present framework, we classify the equations of momentum and render conditions for the Cauchy-Kowalevski theorem to apply. All the above are valid for a simplified model amenable to closed form solutions. For the general problem we lay down the equations of momentum, classify them and give conditions for the Cauchy-Kowalevski theorem to apply. 

As for future directions, we consider that investigation of thin graphene sheets on substrates constitutes a highly challenging theoretical and experimental problem; linearization of the present framework (\cite{Sfyrisetal2014}) will provide system of equations that can be treated easier and give interesting results. More specifically, the linearized equations together with the incorporation of substrate effects to the model, will make the present approach more relevant to actual experimental set-ups such as \cite{Androulidakisetal2014}. \\

\setcounter{section}{0}\setcounter{equation}{0}
\renewcommand{\thesection}{Appendix \Alph{section}}
\renewcommand{\theequation}{\Alph{section}-\arabic{equation}}
\section{Quasilinear first order systems}\label{appen}

This chapter presents a short reminder of some parts of the theory of quasilinear first order systems of partial differential equations based on \cite{Ockendonetal}. Use of the theory of quasilinear systems to the present work consists of viewing the equations of momentum or the equations ruling the shift vector as a quasilinear system in terms of the shift vector, when loading is specified. 

A quasilinear first order system of partial differential equations is defined by (\cite{Ockendonetal})
\begin{equation}
{\bf A} (x, y, {\bf u}) \frac{\partial {\bf u}}{\partial x} +{\bf B}(x, y, {\bf u}) \frac{\partial {\bf u}}{\partial y}= {\bf c}(x, y, {\bf u}).
\end{equation}
The domain where the system is to be solved is defined by the coordinates $x, y$ while the unkwown function is $\bf u$. Cauchy data specify $\bf u$ on a curve $\Gamma$ in the $(x, y)$ plane, i.e. 
\begin{equation}
x=x_0(s), \ \ y=y_0(s), \ \ {\bf u}={\bf u}_0(s). 
\end{equation}
The system of eq. (A-1) together with the data of eq. (A-2) constitutes the Cauchy problem. The condition on the initial data for the first derivative of $\bf u$ to be locally determined is
\begin{equation}
det(x_0' {\bf B}-y_0' {\bf A}) \neq 0.
\end{equation}
This condition may equivalently be given as 
\begin{equation}
\textrm{det} \left(\begin{array}{cccc}
a_{11} & a_{12} & b_{11} & b_{12}       \\
a_{21} & a_{22} & b_{21} & b_{22}       \\
x_0' & 0 & y_0' & 0       \\
0 & x_0' & 0 & y_0'
\end{array}\right) \neq 0
\end{equation}

When $\bf A$ is invertible, the Cauchy-Kowalevski theorem renders existence and uniqueness for an analytic function ${\bf u}={\bf u}(x, y)$ in a neighborhood of $(0, y_0)$ for the system 
\begin{eqnarray}
&&  {\bf u}={\bf u} (y), \ \ \textrm{on} \ \ x=0 \nonumber\\
&& \frac{\partial {\bf u}}{\partial x}={\bf A}^{-1} {\bf c}-{\bf A}^{-1} {\bf B} \frac{\partial {\bf u}}{\partial y}
\end{eqnarray}
provided the right hand side of eq. (A-5)$_2$ is analytic in its arguments. For such a problem the slopes of the characteristics satisfy the eigenvalue problem
\begin{equation}
\frac{d y}{d x}=\lambda, \ \ \ \textrm{where} \ \ \ \textrm{det} ({\bf B}-\lambda {\bf A})=0.
\end{equation}
To classify the system we need to characterize the eigenvalues $\lambda$. When there are two distinct real eigenvalues the system is hyperbolic, when there is a double real eigenvalue the system is parabolic, while when the eigenvalues are complex the system is elliptic. \\

$\bf {Acknowledgements}$ \\

This research has been co-financed by the European Union (European Social Fund - ESF) and Greek national funds through the Operational Program "Education and Lifelong Learning" of the National Strategic Reference Framework (NSRF) - Research Funding Program: ERC-10 "Deformation, Yield and Failure of Graphene and Graphene-based Nanocomposites". The financial support of the European Research Council through the projects ERC AdG 2013 (‘‘Tailor
Graphene’’) is greatfully acknowledged. Valuable discussions with G. Dassios (Patras, Greece) are really appreciated. Special thanks go to E. Koukaras (Patras, Greece) for drawing the figures as well as for his comments regarding the manuscript. Last but not least we would like to thank the reviewers for their time and efforts reviewing towards the improvement of this work.   




\vspace{0.3cm}
D. Sfyris \\
FORTH/ICE-HT, Patras, Greece\\
dsfyris@iceht.forth.gr \\
\vspace{0.3cm}\\
G.I. Sfyris\\
LMS, Ecole Polytechnique, Paris, France\\
\vspace{0.3cm}\\
C. Galiotis \\ 
FORTH/ICE-HT, Patras, Greece and \\
Department of Chemical Engineering, University of Patras, Patras, Greece

\end{document}